\newif\ifconference
\newif\ifanonymous
\conferencefalse
\anonymousfalse

\documentclass[acmsmall,screen,nonacm,natbib=false]{acmart}

\pdfoutput=1

\title{Universal Optimality of Dijkstra via Beyond-Worst-Case Heaps}%

\author{Bernhard Haeupler}
\affiliation{%
	\institution{INSAIT, Sofia University ``St. Kliment Ohridski''}
	\country{Bulgaria}
}
\affiliation{%
	\institution{ETH Zurich}
	\country{Switzerland}
}
\email{bernhard.haeupler@inf.ethz.ch}

\author{Richard Hladík}
\affiliation{%
	\institution{ETH Zurich}
	\country{Switzerland}
}
\affiliation{%
	\institution{INSAIT, Sofia University ``St. Kliment Ohridski''}
	\country{Bulgaria}
}
\email{rihl@uralyx.cz}

\author{Václav Rozhoň}
\affiliation{%
	\institution{Charles University}
	\country{Czech Republic}
}
\affiliation{%
	\institution{INSAIT, Sofia University ``St. Kliment Ohridski''}
	\country{Bulgaria}
}
\email{vaclavrozhon@gmail.com}

\author{Robert E.~Tarjan}
\affiliation{%
	\institution{Princeton University}
	\country{USA}
}
\email{ret@cs.princeton.edu}

\author{Jakub Tětek}
\affiliation{%
	\institution{INSAIT, Sofia University ``St. Kliment Ohridski''}
	\country{Bulgaria}
}
\email{j.tetek@gmail.com}

\newif\ifjacm
\jacmtrue
\makeatletter
\@ifundefined{ifjacm}{%
  \newif\ifjacm
  \jacmfalse
}{}
\makeatother

\ifjacm\else
\usepackage[margin=1in]{geometry}
\usepackage[dvipsnames,svgnames]{xcolor}
\fi
\usepackage{algorithm}
\usepackage{algorithmicx}
\usepackage{algpseudocodex}
\usepackage{xspace}

\usepackage{amsmath}        
\usepackage{amsfonts}       
\usepackage{amsthm}         
\usepackage{thmtools}
\usepackage{thm-restate}
\ifjacm\else
\usepackage{amssymb}
\fi
\usepackage{bbding}         
\usepackage{bm}             
\usepackage{dsfont}
\usepackage{textcomp}
\usepackage{graphicx}       
\usepackage{fancyvrb}       
\usepackage{paralist}
\usepackage{mathtools}

\usepackage[textwidth=20mm,obeyFinal,colorinlistoftodos]{todonotes}

\setuptodonotes{size=\tiny}

\ifjacm
\RequirePackage[ datamodel=acmdatamodel, style=acmauthoryear, backend=biber ]{biblatex}
\else
\ifconference
\let\citet\cite
\else
\usepackage[backend=biber,style=alphabetic,sorting=nty,maxnames=6]{biblatex}         
\let\citet\textcite
\fi
\fi

\usepackage{hyperref}
\ifjacm\else
\hypersetup{
    unicode=true,          
    colorlinks=true,        
    linkcolor=red,          
    citecolor=DarkGreen,        
    filecolor=magenta,      
    urlcolor=cyan           
}
\fi
\usepackage{enumitem}
\usepackage{graphicx} 

\ifconference\else
\usepackage{appendix}
\fi

\usepackage[capitalize, nameinlink,noabbrev]{cleveref}
\usepackage{lineno}

\ifjacm\else
\makeatletter
\AtBeginDocument{
  \hypersetup{
    pdftitle = {\@title},
    pdfauthor = {\ourauthors},
	pdfsubject = {\ourabstract},
	pdfkeywords = {\ourkeywords}
  }
}
\makeatother
\fi

\ifanonymous
\linenumbers
\fi

\renewcommand\O{\mathrm{O}}
\let\OO\O

\newtheorem{theorem}{Theorem}[section]
\newtheorem{lemma}[theorem]{Lemma}
\newtheorem{meta-theorem}[theorem]{Meta-Theorem}

\newtheorem{remark}[theorem]{Remark}
\newtheorem{corollary}[theorem]{Corollary}

\crefname{theorem}{Theorem}{Theorems}
\crefname{proposition}{Proposition}{Propositions}
\crefname{observation}{Observation}{Observations}
\crefname{lemma}{Lemma}{Lemmas}
\crefname{claim}{Claim}{Claims}
\crefname{problem}{Problem}{Problems}
\crefname{conjecture}{Conjecture}{Conjectures}
\crefname{question}{Question}{Questions}
\crefname{example}{Example}{Examples}
\crefname{fact}{Fact}{Facts}    
\crefname{invariant}{Invariant}{Invariants}

\newcounter{Halgorithmic}
\AtBeginEnvironment{algorithmic}{\stepcounter{Halgorithmic}}
\ExpandArgs{c}\renewcommand{theHALG@line}{\arabic{Halgorithmic}.\arabic{ALG@line}}

\let\cref=\Cref
\crefname{lemma}{Lemma}{Lemmas}
\crefname{theorem}{Theorem}{Theorems}

\newcommand{\el}{\ell}

\newcommand{\Null}{\mathit{null}}

\newcommand{\Dijkstra}{\textsc{Dijkstra}}
\newcommand{\Scan}{\textsc{Scan}}

\ifjacm
\newcommand{\Next}{\textit{next}}
\newcommand{\Prev}{\textit{prev}}
\newcommand{\MakeSet}{\textit{make-set}}
\newcommand{\Find}{\textit{find}}
\newcommand{\Unite}{\textit{unite}}
\newcommand{\MakeHeap}{\textit{make-heap}}
\newcommand{\Insert}{\textit{insert}}

\newcommand{\DeleteMin}{\textit{delete-min}}

\newcommand{\FindMin}{\textit{find-min}}
\newcommand{\DecreaseKey}{\textit{decrease-key}}
\newcommand{\Meld}{\textit{meld}}
\else
\newcommand{\Next}{\textsc{Next}}
\newcommand{\Prev}{\textsc{Prev}}
\newcommand{\MakeSet}{\textsc{MakeSet}}
\newcommand{\Find}{\textsc{Find}}
\newcommand{\Unite}{\textsc{Unite}}
\newcommand{\MakeHeap}{\textsc{MakeHeap}}
\newcommand{\Insert}{\textsc{Insert}}

\newcommand{\DeleteMin}{\textsc{DeleteMin}}

\newcommand{\FindMin}{\textsc{FindMin}}
\newcommand{\DecreaseKey}{\textsc{DecreaseKey}}
\newcommand{\Meld}{\textsc{Meld}}
\fi

\newif\ifmeld\meldfalse

\ifanonymous
\def\ourauthors{Anonymous authors}
\else
\def\ourauthors{Bernhard Haeupler, Richard Hladík, Václav Rozhoň, Robert E. Tarjan, Jakub Tětek}
\fi

\def\ourabstract{%
In this paper we prove that Dijkstra's shortest-path algorithm, if implemented with a sufficiently efficient heap, is universally optimal in its running time, and with suitable small additions is also universally optimal in its number of comparisons.
\texorpdfstring{\par}{}
Universal optimality is a powerful beyond-worst-case performance guarantee for graph algorithms that informally states that a single algorithm on a problem involving graphs with arc and/or vertex weights performs as well as possible on \emph{every} graph, assuming a worst-case choice of weights. 
We give the first application of this notion to any sequential algorithm. 
\texorpdfstring{\par}{}
We design a new heap data structure with a \emph{working-set bound}, which guarantees that the heap takes advantage of a certain kind of locality in the heap operations. Our heap has the optimal (amortized) bounds of Fibonacci heaps but also has the beyond-worst-case guarantee that the cost of deleting the minimum item is logarithmic in the number of items inserted after it but before it is deleted, instead of logarithmic in the size of the heap when the item is deleted. That is, deletion of recently inserted items is especially efficient.  
\texorpdfstring{\par}{}
We prove that our working-set bound guarantees universal optimality for the problem of ordering vertices by their distance from the source vertex, which we call the \emph{distance order problem}.  Our result relies on the observation that the sequence of heap operations generated by any run of Dijkstra's algorithm on a fixed graph possesses enough locality that one can couple the number of comparisons performed by any heap with our working-set bound to the minimum number of comparisons required to solve the distance order problem on this graph for a worst-case choice of arc lengths. 
}

\def\ourkeywords{Dijkstra's algorithm, universal optimality, beyond-worst-case heaps, working-set bound, shortest paths}

\begin{document}
\let\coloneq\coloneqq


\newpage

\ifjacm\else
\maketitle
\fi

\begin{abstract}
\ourabstract
\end{abstract}


\ifjacm
\maketitle
\fi

\hypersetup{
pdfsubject = {We prove that Dijkstra's shortest-path algorithm, if implemented with a sufficiently efficient heap, is universally optimal in its running time, and with suitable small additions is also universally optimal in its number of comparisons.}
}

 \thispagestyle{empty}
\newpage
\clearpage
\tableofcontents

 \thispagestyle{empty}
\newpage
\clearpage
\setcounter{page}{1}

\section{Introduction}

Universal optimality is a powerful beyond-worst-case performance guarantee for algorithms on weighted graphs.  Informally, it states that a single algorithm runs as fast as possible on any graph, for a worst-case choice of arc weights.

This paper gives the first application of this notion to the standard sequential model of computation.  We prove that Dijkstra's shortest path algorithm, when implemented using a sufficiently efficient heap data structure and with suitable small additions, is universally optimal for the natural problem of ordering the vertices by their distance from the source.

We also design a heap that has the needed efficiency.  Our new heap is a strict improvement over the Fibonacci heap~\cite{fredman1987fibonacci} and similar data structures.  It has a beyond-worst-case bound for delete-min operations while preserving the $\OO(1)$ time bound for insertion and decrease-key operations.

Specifically, we show that a natural \emph{working-set bound} for heaps, related to one originally proposed by~\citet{iacono2000improved}, suffices to make Dijkstra's algorithm universally optimal with respect to running time.  This bound states that the time to do a delete-min of an item $x$ is at most logarithmic in the number of items inserted into the heap from the time $x$ is inserted until the time it is deleted: Items inserted before $x$ do not count in the deletion time of $x$.

Our universal optimality result reveals a surprisingly clean connection between the working-set bound and the efficiency of Dijkstra's algorithm: A heap with this bound enables the algorithm to leverage the structure of the problem graph to the fullest extent that is possible with a comparison-based algorithm.

In addition, we present two extensions of Dijkstra's algorithm that are universally optimal with respect to both running time and number of comparisons of sums of arc weights.  Although the time complexity of Dijkstra's algorithm for any graph with $n$ vertices and $m$ arcs lies in the relatively narrow range of $\Omega(m+n)$ to $\OO(m+n\log n)$, the number of comparisons needed to order the vertices by distance from the source can be as few as zero.

Beyond our results on Dijkstra's algorithm, we hope that our paper opens doors for future research on applying variants of universal optimality to problems in the standard sequential model of computation.

\subsection{Beyond the worst case: universal optimality}

The notion of asymptotic worst-case complexity is a foundational concept of theoretical computer science and algorithm design.  Discoveries by many researchers over many decades have produced for more and more problems state-of-the-art algorithms whose efficiency is essentially best possible in the worst case. A worst-case guarantee may not be completely satisfactory, however: Just because there is some family of instances on which an algorithm cannot perform well does not mean one should be satisfied with an algorithm that performs badly on easy instances as well.

One approach to addressing this issue is the notion of parameterized complexity, in which each instance $x$ of a problem not only has a size $|x|$ but also a complexity parameter $\kappa(x)$, and efficiency bounds for algorithms are stated as functions of both $|x|$ and $\kappa(x)$.  Ideally, one can obtain an algorithm that is optimal with respect to both $|x|$ and $\kappa(x)$ -- that is, no correct algorithm can have better complexity with respect to this parameterization.

Taking parameterized complexity to its extreme, one may choose $\kappa(x) = x$; that is, parameterize by the instance itself.  This results in the notion of \emph{instance optimality}~\cite{fagin2001first_instance_optimal}.  An instance-optimal algorithm is at least as efficient as any correct algorithm, on every single input.

Sadly, instance-optimal algorithms are rare.  For algorithms on weighted graphs, however, there is a more realistic alternative to instance optimality: \emph{universal optimality}~\cite{haeupler2021universally}.  The idea is to parametrize an instance by the graph but not the weights.  That is, an algorithm $A$ is universally optimal if, on any particular graph $G$ and for any other algorithm $A'$ that is correct on $G$, the efficiency of $A$ on $G$ for a worst-case choice of weights is asymptotically as good as the efficiency of $A'$ on $G$ for a worst-case choice of weights.

Intuitively, a universally optimal algorithm is as efficient as possible on any given graph.  If for example there are fast algorithms for planar graphs, then a universally optimal algorithm must be fast when run on any planar graph.  This must be true for any subclass of graphs and indeed for any specific graph.

\subsection{An illustrative example}

To begin to address the question of whether Dijkstra's algorithm is universally optimal, and, if not, how to make it so, let us consider the graph containing $n=r+t+1$ vertices and $m=r+t$ arcs, whose vertices are the source $s$, a vertex $u_i$ for $1 \leq i \leq t$, and a vertex $v_j$ for $1 \leq j \leq r$, and whose arcs are $sv_1$, $su_i$ for $1 \leq i \leq t$, and $v_jv_{j+1}$ for $1\leq j<r$.  (See~\cref{fig:broom}.)

\begin{figure}
    \centering
    \includegraphics[width=\ifconference\hsize\else.7\hsize\fi]{fig/broom.pdf}
    \caption{An example graph. Dijkstra's algorithm implemented with a Fibonacci heap needs $\Omega(r \log t)$ time to finish. For $t \ll r$, this is not optimal, as there exists an $\O(r + t \log t)$-time algorithm. }
	\Description{Fully described in the text.}
    \label{fig:broom}
\end{figure}

Given a directed graph such as our example and given that each arc $vw$ has a non-negative length $c(vw)$, Dijkstra's algorithm computes the length of a shortest path from a given source vertex, in our example $s$, to each vertex reachable from $s$.  It does this using the greedy method.  For each vertex $v$, it maintains a \emph{current distance} $d(v)$ equal to the shortest length of a path from $s$ to $v$ found so far, initially $0$ for $s$ itself, $\infty$ for every other vertex.  It maintains a partition of the vertices into three sets: \emph{unlabeled, labeled, and scanned}.  The unlabeled vertices are those with infinite current distance.  Each vertex with finite current distance is either labeled or scanned.  Initially $s$ is labeled and no vertex is scanned.  The algorithm repeats the following step until there are no labeled vertices: Choose a labeled vertex $v$ with minimum current distance, make $v$ scanned, and for each arc $vw$, if $w$ is unlabeled, set $d(w)\gets d(v)+c(vw)$ and make $v$ labeled; otherwise, if $w$ is labeled and $d(w) > d(v)+c(vw)$, set $d(w)\gets d(v)+c(vw)$.

The non-negativity of arc lengths implies that when a vertex becomes scanned, its current distance is its \emph{true distance}; namely, the shortest length of a path from $s$ to $v$.  Finally, vertices become scanned in non-decreasing order by their distance from $s$.

Efficient implementations of Dijkstra's algorithm store the set of labeled vertices in a heap, with the key of a vertex equal to its current distance.  If the problem graph has $n$ vertices and $m$ arcs, there are $n$ heap insertions, at most $n$ delete-min operations, and at most $m-n+1$ decrease-key operations.  If the heap is a Fibonacci heap or a comparably efficient heap, the worst-case running time of the algorithm is $\OO(m+n\log n)$: Each of the insertions and decrease-key operations takes $\OO(1)$ amortized time, and each of the delete operations takes $\OO(\log n_i)$ amortized time, where $n_i$ is the number of labeled vertices when the delete-min occurs.

Suppose Dijkstra's algorithm is run on our example graph using a heap, and suppose that the length of the path from $s$ to $v_r$ is smaller than the length of every arc $su_i$.  Then the heap deletions will be of $s$, then the $v_j$ in increasing order by $j$, and then the $u_i$ in some order.  When $v_j$ is deleted, the heap size is $t+1$.  If the heap is a Fibonacci heap or any similar kind of heap, and $t$ is chosen appropriately, the time for each deletion of a vertex $v_j$ will be $\Omega(\log t)$, resulting in a total time of $\Omega(r\log t)$.  (Verifying that this lower bound holds for each particular kind of heap requires examining the details of how the heap works and choosing $t$ carefully, generally a power of two.)  On the other hand, one can compute distances from $s$ and sort the vertices by distance as follows: Sort the arc lengths of $su_i$ for $1 \leq i \leq t$; compute the true distance of each $v_j$ from $s$ by summing arc lengths along the path from $s$ to $v_r$; merge in sorted order by distance the list of $u_i$ sorted by distance with the list of $v_j$ sorted by index.  The total time is $\OO(r+t\log t)$, which is asymptotically smaller than $r\log t$ if for example $r=t^2$.  Thus Dijkstra's algorithm implemented with a Fibonacci (or similar) heap is \emph{not} optimal on this graph.

Observe, though, that every vertex $u_i$ is inserted into the heap \emph{before} $v_j$ for $j>1$.  Let us define the \emph{working set} of a vertex $v$ to be the number of vertices inserted into the heap between the time $v$ is inserted and the time it is deleted, including $v$.  Then the working-set size of each $v_j$ is just~$1$.  If the time to delete a vertex from the heap is logarithmic in the size of its working set, then Dijkstra's algorithm will run in $\OO(r+t\log t)$ time on our example graph.  Fulfilling the promise of this observation is the goal of our paper.

\subsection{Roadmap}

This paper is a substantially rewritten and improved version of our conference paper \cite{dijkstra-focs}.  It contains ten sections in addition to this introduction.  \cref{sec:related-work} briefly reviews related work, including work on shortest paths for graphs with non-negative arc lengths, on universal and instance optimality, on working-set bounds and other beyond-worst-case bounds for data structures, and on sorting using partial information, a problem to which our ideas also apply. \cref{sec:graph-concepts} formally defines the graph concepts we use, including those related to shortest paths, in particular the definition of the main problem we consider, the distance order problem.  It also presents concepts related to unavoidable vertices on paths from the source. \cref{sec:dijkstra} gives a high-level description of Dijkstra's algorithm and its implementation using a heap.  It includes a formal definition of the working-set bound for heaps. \cref{sec:complexity-models} formally defines the two complexity models we consider, which measure time and comparison complexity, respectively, and it defines universal optimality in these two models.  \cref{sec:lower-bounds} derives lower bounds for the time complexity and comparison complexity of the distance order problem.  

\cref{sec:dijkstra-efficiency} proves that Dijkstra's algorithm using a heap with the working-set bound runs in time matching the lower bound on time proved in~\cref{sec:lower-bounds}, thus showing that this algorithm is universally optimal with respect to time, and universally optimal in comparisons up to an additive term linear in the number of vertices.  To obtain this result, we consider the \emph{search tree} generated by a run of Dijkstra's algorithm.  This is the tree $T$ rooted at the source vertex $s$ containing each arc $vw$ whose processing during the run of the algorithm makes $w$ labeled.  \cref{sec:lower-bounds} includes a proof that any correct algorithm for the distance order problem must do $\Omega(\log D(T))$ comparisons, where $D(T)$ is the number of topological orders of $T$.  We relate this bound to the sum of the logarithms of the working-set sizes of the vertices deleted from the heap during the run of Dijkstra.  As a bridging concept, we define an interval graph based on the intervals during which vertices are in the heap.  We use a lemma of~\citet{vanderhoog2024} to bound the number of topological orders of this interval graph, which is both a lower bound on $D(T)$ and (to within a constant factor) an upper bound on the product of the working-set sizes.  This result is one of the main technical contributions of the paper.  The use of the lemma of~\cite{vanderhoog2024} allows us to significantly simplify the proof in the conference version of our paper~\cite{dijkstra-focs}.

\cref{sec:dijkstra-lookahead} extends Dijkstra's algorithm to make it universally optimal in both time and comparisons.  This requires reducing the number of comparisons done by the algorithm if the required number is sublinear in the number of vertices.  To do this we use \emph{bottlenecks}.  Assign to each vertex~$v$ an integer level equal to the minimum number of vertices on the path from the source $s$ to $v$.  A vertex is a bottleneck if it is the only vertex on its level.  If $v$ is a bottleneck, every path from $s$ to any vertex on a level higher than that of $v$ must contain $v$.  We add a preprocessing step to Dijkstra's algorithm that finds all the bottlenecks, which takes linear time and no comparisons using breadth-first search.  Then we run Dijkstra's algorithm, but without inserting the bottlenecks into the heap.  Instead, the algorithm computes the true distances for the bottlenecks proactively, and adds the bottlenecks to the distance order as vertices at greater distances are deleted from the heap.  We call the resulting algorithm \emph{Dijkstra with lookahead}.  It is substantially simpler than the corresponding algorithm in the conference version of our paper~\cite{dijkstra-focs}.

\cref{sec:dijkstra-recursive} presents a different way to make Dijkstra's algorithm universally optimal in time and comparisons, by calling it recursively.  Like the algorithm in \cref{sec:dijkstra-lookahead}, the recursive algorithm computes bottlenecks, but it uses them differently.  It runs Dijkstra's algorithm from the original source.  When it scans a bottleneck, it begins a new, recursive run of Dijkstra's algorithm from this bottleneck.  Once the recursive run is finished, the original run resumes and runs to completion.  We call this algorithm \emph{recursive Dijkstra}.  Recursive Dijkstra does not necessarily scan the vertices in order by distance, so to solve the distance order problem we need an additional data structure to maintain the scanned vertices in distance order.  A homogeneous finger search tree \cite{HuddlestonM82} suffices.  

\cref{sec:heap} develops a heap with the working-set bound.  If the only required heap operations are \Insert{} and \DeleteMin{}, then a pairing heap suffices, as proved by~\citet{iacono2000improved}.  But in addition we need to support \DecreaseKey{} operations in $\OO(1)$ amortized time, which requires a new data structure.  Our structure consists of a list of \emph{inner heaps}, each of which is a Fibonacci heap or equivalently efficient heap.  We obtain the working-set bound by making sure that each item in an inner heap earlier in the list was inserted after all items in inner heaps later in the list.  Thus each item in an inner heap is in the working set of all items in inner heaps later in the list.  Roughly speaking, we maintain the inner heap sizes so that they grow doubly exponentially.  This gives us the working-set bound for \DeleteMin{} operations, and it limits the number of inner heaps to be doubly logarithmic.  We need to keep track of a set of minimum-key items, one per inner heap, but since the number of inner heaps is very small we can use bit-vector techniques to do this.  The heap we describe here is somewhat simpler than the one in the conference version of our paper~\cite{dijkstra-focs}

Lastly, \cref{sec:remarks} contains some final remarks.

\section{Related Work}\label{sec:related-work}

\subsection{Shortest path algorithms}

Dijkstra's algorithm is a foundational method for solving the single-source shortest path problem with non-negative arc lengths.  Dijkstra's original implementation runs in $\OO(n^2)$ time on an $n$-vertex graph.  For sparse graphs (in which $m$, the number of arcs, is much smaller than $n^2$), the running time can be reduced to $\OO(m\log n)$ by implementing the algorithm using a classical heap such as that of~\citet{Williams64}. \citet{fredman1987fibonacci} reduced the time bound to $\OO(m+n \log n)$ by using a Fibonacci heap, which they invented for this purpose and for other applications.  This running time is best possible as a function of $n$ and $m$ if we require the algorithm to produce a sorted order of the vertices by distance from the source, which Dijkstra's algorithm does.  For the problem of finding the distances without sorting them, the recent breakthrough result of~\citet{duan2025breakingsortingbarrierdirected} runs in $\OO(m\log^{2/3} n)$ time, thus showing that Dijkstra's algorithm is not optimal for this problem.

If the arc lengths are integers, or the ratio between the maximum and minimum arc weight is appropriately bounded, there are many algorithms with a time bound very close to linear~\cite{FW93,FW94, Thorup96, Raman96, Raman97, TM00, HT20,Thorup00,Thorup04}.

\subsection{Universal and instance optimality}

Researchers in the field of distributed algorithms coined the term ``universal optimality."  They developed techniques for designing distributed algorithms that are close to universally optimal for many problems, notably approximate shortest path problems~\cite{ghaffari2021universally,goranci2022universally,haeupler2021universally,haeupler2022hop,rozhon_grunau_haeupler_zuzic_li2022deterministic_sssp}. As far as we know, our extension of the concept of universal optimality to sequential graph algorithms is new.

The notion of universal optimality is a weakening of that of \emph{instance optimality}~\cite{fagin2001first_instance_optimal,valiant2017automatic,chan2017instance_optimal_hull}, which is the gold standard for beyond-worst-case analysis.  Indeed, in the influential book by \citet{roughgarden_2021}, which contains 30 chapters discussing various approaches to beyond-worst-case analysis, instance optimality is covered in the third chapter, after only the introduction and a chapter on parameterized algorithms.  Unfortunately, instance optimality is an extremely strong requirement and thus very hard to satisfy.  In particular, instance optimality is not achievable for our distance order problem, because on a particular instance of a graph with arc lengths, the instance-specific algorithm that merely chooses a correct distance order and then verifies it takes only linear time.  Consequently, results proving instance optimality both restrict the computation model (to allow the derivation of lower bounds) and relax the notion of instance optimality (to disallow uninteresting counterexamples).

One example of this is the work of~\citet{chan2017instance_optimal_hull}. They develop algorithms for a number of geometric problems, including the two-dimensional convex hull problem.  Their computational model is a suitable generalization of the standard comparison model.  They achieve order-oblivious instance optimality, a relaxed version of instance optimality.  One can view the convex hull problem as an extension of the standard sorting problem.  Analogously, our distance order problem can be viewed as an extension of sorting.  There may well be other ordering problems for which instance-optimal or universally optimal algorithms can be obtained.

\subsection{Beyond-worst-case bounds for data structures}

A famous open problem in data structures, the dynamic optimality conjecture for splay trees~\cite{sleator1985self,iacono2013pursuit}, is a question of instance optimality.  A related question for heaps fails for a large class of heaps known as tournament heaps~\cite{munro2019dynamic}.

Splay trees are known to have a number of beyond-worst-case bounds, including a so-called ``working-set" bound.  \citet{iacono2000improved} initiated the study of similar bounds for heaps.   He showed that the pairing heap, a well-known self-adjusting heap~\cite{fredman1986pairing}, has a working-set bound if \DecreaseKey{} is not a supported operation.  We use a working-set bound whose definition is apparently weaker than Iacono's but that (in so-far-unpublished work) we have proved to be asymptotically equivalent to Iacono's. \citet{elmasry} and \citet{elmasry_farzan_iacono} gave heap implementations that have stronger working-set-type bounds, but again assuming that \DecreaseKey{} is not supported.  The heap used in the implementation of Dijkstra's algorithm needs to support many \DecreaseKey{} operations, so these results do not help us here.

\subsection{Sorting using partial information} After publicizing a preliminary version of this paper, the authors together with John Iacono applied some of its ideas to develop an algorithm for the problem of sorting a set of numbers given the outcomes of some pre-existing comparisons.  The paper~\cite{supi} presents this algorithm and related results.  The problem itself dates to a conjecture of~\citet{kislitsin} and to a classical paper of~\citet{fredman-generalized-supi-1976}.  One can represent the set of pre-existing comparison outcomes by a directed acyclic graph (DAG).  The problem then becomes that of finding the unknown total order of the vertices by doing additional comparisons.  The algorithm of~\citet{supi} does $\OO(\log T)$ comparisons and runs in $\OO(m+n+\log T)$ time, given an input DAG with $n$ vertices, $m$ arcs, and $T$ topological orders.  These bounds are best possible.

The present work and~\cite{supi} have many parallels.  The basic algorithm in~\cite{supi}, \emph{topological heapsort}, can be viewed as a variant of Dikstra's algorithm, and the comparison-optimal algorithm, \emph{topological heapsort with lookahead}, is a variant of an algorithm we develop here, \emph{Dijkstra with lookahead}.  The proof that topological heapsort is universally optimal uses our framework for universal optimality as well as some of the analytic techniques we use here.  

\section{Shortest Paths and Other Graph Concepts}\label{sec:graph-concepts}

Throughout this paper $G$ is a directed graph with a specified \emph{source vertex} $s$ such that all vertices are reachable from $s$.  We denote by $vw$ an arc from vertex $v$ to vertex $w$.  Except where otherwise noted, we assume that $G$ contains no multiple arcs and no self-loops, so our notation for arcs is unambiguous.  We can represent an undirected graph as a directed graph by replacing each edge connecting two vertices $v$ and $w$ by the pair of arcs $vw$ and $wv$, each with a length equal to the length of the replaced edge.  Our results on the shortest path problem also hold for undirected graphs, as we discuss result by result.  For an undirected graph, we denote the number of edges by $m$.  Hence the corresponding directed graph contains $2m$ arcs.

We denote by $n$ and $m$ the number of vertices and arcs of $G$, respectively.  To simplify the statement of certain bounds we assume $n >2$.   This assumption implies $n= \OO(m)$, since $m \geq n - 1 \geq 2n/3$.

Suppose each arc $vw$ in $G$ has a non-negative real-valued \emph{length} $c(vw)$.  The \emph{length} of a path $P$ in $G$ is the sum of its arc lengths.  A path from $v$ to $w$ is \emph{shortest} if it has minimum length among all paths from $v$ to $w$.  Since all arc lengths are non-negative, if there is any path from $v$ to $w$ there is a shortest \emph{simple} path from $v$ to $w$, i.e., a path that repeats no vertex.  We denote by $d^*(v)$ the length of a shortest path from $s$ to $v$.  We call $d^*(v)$ the \emph{true distance} from $s$ to $v$.  A \emph{shortest path tree} of $G$ is a spanning tree rooted at $s$ all of whose paths are shortest.

In studying shortest path algorithms we shall need some additional concepts.  A \emph{topological order} of a directed acyclic graph is a total order of the vertices such that if $vw$ is an arc, $v$ is less than $w$ in the order. A \emph{distance order} of an arbitrary directed graph $G$ with source $s$ is a total order of the vertices having the property that there is some assignment of non-negative lengths to the arcs such that all true distances from the source are distinct, and the order is strictly increasing by true distance.  We denote a distance order by a list $L$ of the vertices in the given order.  We denote by $D$ the number of distance orders of the problem graph $G$ with source vertex $s$.  The following lemma gives a useful characterization of distance orders:

\begin{lemma}\label{lem:tree-top-dist}  A total order of the vertices of $G$ is a distance order if and only if for every vertex $w\neq s$, there is an arc $vw$ such that $v$ precedes $w$ in $L$.
\end{lemma}
\begin{proof}
Suppose $L$ is a distance order.  Consider a set of non-negative arc lengths such that the true distances are distinct and $L$ is ordered by increasing true distance.  If $w\neq s$, let $vw$ be the last arc on a shortest path from $s$ to $w$.  Then $d^*(w)=d^*(v)+c(vw)$.  Since true distances are distinct and arc lengths are non-negative, $d(w)>d(v)$, so $v$ precedes $w$ in $L$, as required.

Conversely, suppose $L=[v_1, v_2,\dots, v_n]$ be a total order of the vertices such that if $w\neq s$ there is an arc $vw$ with $v$ preceding $w$ in the order.  Then $v_1=s$.  For each vertex $v_j\neq s$, choose an arc $v_iv_j$ such that $i < j$ and set its length equal to $j-i$.  Set the length of every other arc equal to $n$.  Then the true distance of $v_i$ from $s$ is $i-1$.  Hence $L$ is a distance order.
\end{proof}

\begin{remark}
In the case of an undirected graph, each arc becomes two oppositely directed arcs, each with the same length.  The proof of~\cref{lem:tree-top-dist} holds for such graphs if each arc $v_jv_i$ with $i< j$ is given the same length as $v_iv_j$.
\end{remark}

Given a total order of the vertices, an arc $vw$ is a \emph{forward arc} of the order if $v$ precedes $w$ in the order.  Thus a total order of the vertices is a distance order if and only if each vertex other than $s$ has an incoming forward arc. We denote by $F$ the number of forward arcs of a distance order whose number of forward arcs is maximum.  For every distance order, there are at least $n-1$ forward arcs, those of a spanning tree for which the distance order is a topological order.  Since we are assuming $n>2$, $n=\OO(F)$.  In the case of an undirected graph, each undirected edge becomes two directed arcs, exactly one of which is forward.  Hence for undirected graphs $F=m$, where $m$ is the number of edges in the undirected graph. 

Given a set of arc lengths, a \emph{true distance order} is a distance order that is non-decreasing with respect to the true distances for the given arc lengths.  That is, the vertices are in non-decreasing order by true distance, and every vertex $v\neq s$ has an incoming forward arc.  The main problem we consider is that of computing a true distance order of a graph $G$.

\begin{remark}
Our definition of a true distance order is designed to allow arcs to have length zero.  Dijkstra's algorithm is correct for arbitrary non-negative arc lengths, and we want results that encompass this generality.  If we did not impose the requirement on a distance order that each vertex other than $s$ has an incoming forward arc, then in a graph with \emph{all} arc lengths zero, \emph{any} order of the vertices would be a non-decreasing order by true distance, which is not what we want.   
\end{remark}

Our final graph concepts involve unavoidable vertices on paths from $s$, specifically \emph{dominators} and \emph{bottlenecks}.

Given two distinct vertices $v$ and $w$, we say $v$ \emph{dominates} $w$ if $v$ is on every path from $s$ to $w$.  The \emph{immediate dominator} of a vertex $v \neq s$ is the unique dominator of $v$ that is dominated by all other dominators of $v$.  Every vertex has an immediate dominator, and the arcs to vertices from their immediate dominators form a tree called the \emph{dominator tree}.  The dominator tree contains all the vertices in $G$, but in general it is not a spanning tree, since its arcs are not necessarily in $G$.  The dominators of a vertex are exactly its proper ancestors in the dominator tree.  For discussions of dominators and proofs of these facts see~\cite{books/lib/AhoU73,journals/cacm/LowryM69}.

An arc $vw$ is \emph{useless} if $w$ dominates $v$ and \emph{useful} otherwise.  A useless arc cannot be a forward arc of any distance order.  Nor can it be on a simple shortest path from $s$ to any vertex, so useless arcs can be ignored when computing shortest paths from $s$.  The set of useless arcs can be found by computing the dominator tree and marking as useless each arc $vw$ such that $w$ is an ancestor of $v$ in the dominator tree.  Computing the dominator tree takes $\OO(m)$ time~\cite{harel1985domtree_linear,alstrup1999domtree_linear,journals/siamcomp/BuchsbaumGKRTW08,bob-dominators-2}.  Testing the ancestor-descendant relation in a static tree takes $\OO(1)$ time per query \cite{journals/siamcomp/Tarjan74}.  Thus the set of useless arcs can be found in $\OO(m)$ time.

Dijkstra's algorithm in fact ignores useless arcs, so our universally comparison-optimal algorithm does not need to find useless arcs explicitly, nor to compute dominators.  It relies instead on a weaker notion, that of \emph{bottlenecks}.  We define the \emph{level} $\ell(v)$ of a vertex $v$ to be the minimum number of vertices on a path from $s$ to $v$.  This definition implies $\ell(v)+1 \geq \ell(w)$ for every arc $vw$.  A vertex $v$ is a \emph{bottleneck} if it is the only vertex on its level.  If $v$ is a bottleneck, it dominates all vertices on higher levels.  The bottlenecks lie on a single path in the dominator tree.  In general not all the vertices on this path are bottlenecks.  We can compute levels and find bottlenecks in $\OO(m)$ time by doing a breadth-first search from $s$.

\section{Dijkstra's Algorithm}\label{sec:dijkstra}

\subsection{A high-level view}\label{sec:dijkstra-high-level}

Here is a complete high-level formal description of Dijkstra's algorithm~\cite{Dij59}.  Given a graph $G$ with a source $s$ and a set of non-negative arc lengths, the algorithm computes the true distance $d^*(v)$ for every vertex $v$ using the greedy method.  The input to the algorithm is the set of vertices in $G$; the source vertex $s$; and, for each vertex $v$, an \emph{incidence list} of its exiting arcs, in no particular order. Each arc stores its length.  The algorithm maintains for each vertex $v$ a \emph{current distance} $d(v)$ equal to the smallest length of a path from $s$ to $v$ found so far, or $\infty$ if no path from $s$ to $v$ has yet been found.  Initially $d(s)=0$ and $d(v)=\infty$ for $v\neq s$.  During a run of the algorithm, each vertex is in one of three states: \emph{unlabeled}, \emph{labeled}, or \emph{scanned}.  Initially $s$ is labeled and all other vertices are unlabeled.  The algorithm initializes the current distances and the states of all vertices and then repeats the following \emph{scanning step} until all vertices are scanned:

\begin{itemize}
    \item [] Choose a labeled vertex $v$ with $d(v)$ minimum.  \emph{Scan} $v$ as follows:  Mark $v$ scanned.  For each arc $vw$, apply the appropriate one (if any) of the following two cases: If  $w$ is unlabeled, set $d(w) \leftarrow d(v)+ c(vw)$ and mark $w$ labeled; if $w$ is labeled and $d(v)+c(vw) < d(w)$, set $d(w) \leftarrow d(v)+ c(vw)$.
\end{itemize}

During a run of the algorithm, the unlabeled vertices are exactly those with current distance infinity.  An induction on the number of scanning steps shows that if $v$ is scanned and $w$ is labeled, $d(v)\leq d(w)$, and that when $v$ becomes scanned, $d(v)=d^*(v)$~\cite{Tarjan1983}.

Dijkstra's algorithm does more than compute the true distances of vertices from $s$.  It builds a shortest path tree, if it is augmented to maintain for each vertex $w\neq s$ an incoming \emph{parent arc} $vw$.  To maintain parent arcs, it sets the parent arc of $w$ equal to $vw$ whenever the processing of an arc $vw$ decreases $d(w)$. When the algorithm stops, the parent arcs form a shortest path tree~\cite{Tarjan1983}.  Also, it scans the vertices in non-decreasing order by true distance~\cite{Tarjan1983}.

\begin{lemma}\label{lem:Dijkstra-scanning-order}
Dijkstras's algorithm scans the vertices in a true distance order.
\end{lemma}
\begin{proof}
Since Dijkstra's algorithm scans vertices in non-decreasing order by true distance, all we need to prove is that scanning order is a distance order.  For any vertex $w\neq s$, $w$ is scanned after it becomes labeled.  For it to become labeled, some vertex $v$ with an arc $vw$ to $w$ must be scanned.  The existence of such an arc $vw$ for each $w\neq v$ implies by~\cref{lem:tree-top-dist} that scanning order is a distance order.   
\end{proof}

Summarizing, Dijkstra's algorithm solves three problems: It computes true distances from the source, it finds a shortest path tree, and it finds a true distance order.  We shall prove that when implemented appropriately, Dijkstra's algorithm finds a true distance order in universally optimal time to within a constant factor, and that the algorithm can be extended to find a true distance order in a universally optimal number of comparisons to within a constant factor.

Let's consider the relationship among the three problems.  Up to constant factors, the distance order problem is the hardest of the three.  We can show this as follows: Suppose we are given a true distance order.  In $\OO(m)$ time and no comparisons we can determine the set of forward arcs for the given order.  Given the set of forward arcs, we can compute the true distances and a shortest path tree by initializing $d(s)\gets 0$ and $d(v)\gets\infty$ for $v \neq s$, initializing $s$ to be labeled and all other vertices to be unlabeled, and then processing the vertices in true distance order.  To process a vertex $v$, for each forward arc $vw$, if $w$ is unlabeled, or $w$ is labeled and $d(w) < d(v)+c(vw)$, set $d(w)\gets d(v)+c(vw)$ and set the parent arc of $w$ equal to $vw$.  After all vertices are processed, the current distances are the true distances, and the parent arcs form a shortest path tree.  This computation takes $\OO(m)$ time and $\OO(F-n+1)$ comparisons, where $F$ is the maximum number of forward arcs of any distance order of the problem graph.  We shall prove in~\cref{sec:lower-bounds} that computing a true distance order takes $\Omega(m)$ time and $\Omega(F-n+1)$ comparisons.  This result implies that the extra time to find the true distances and a shortest path tree is at most a constant factor times the time required to find a true distance order, and the same is true of comparisons.

The problem of computing true distances and that of finding a shortest path tree are closely related.  Given a shortest path tree, one can compute true distances in $\OO(n)$ additions and no comparisons by initializing the true distance of $s$ to be $0$ and summing arc lengths along paths in the tree, proceeding from the root to the leaves.  Given the true distances, one can find a shortest path tree by searching forward from $s$ and when visiting a vertex $v$ adding an arc $vw$ to the tree if $d^*(v)+c(vw)=c^*(w)$ and $w$ does not yet have an incoming tree arc.  This computation takes $\O(m)$ additions and equality tests.


We emphasize that the tight bounds we derive here are for the distance order problem.  It seems much more challenging to obtain a tight bound for either of the other problems, given the recent paper of~\citet{duan2025breakingsortingbarrierdirected} that presents a deterministic algorithm for finding true distances that is asymptotically faster than Dijkstra's algorithm.  This result implies that the distance order problem is asympotically harder than the other two problems in the worst case.


\subsection{Implementation using a heap}\label{sec:dijkstra-heap}

Dijkstra's algorithm reduces the shortest path problem to a data structure problem, that of efficiently maintaining a \emph{heap}.  A heap is a set of items, each having an associated key that is an element in a totally ordered universe.  Heaps support some or all of the following operations:

\begin{itemize}
    
    \item [] $\MakeHeap$: Create and return a new, empty heap.

    \item [] $\FindMin(H)$: Return an item of minimum key in heap $H$, or $\Null$ if $H$ is empty.

	\item [] $\Insert(x, H)$: Insert item $x$, with predefined key, into heap $H$.  Item $x$ must be in no heap before the insertion.

    \item [] $\DeleteMin(H)$: If $H$ is nonempty, delete $\FindMin(H)$ from $H$ and return it; if $H$ is empty, do nothing.

    \item [] $\Meld(H_1, H_2)$: Form and return a heap containing all items in item-disjoint heaps $H_1$ and $H_2$.  The meld destroys $H_1$ and $H_2$.

    \item [] $\DecreaseKey(x, k, H)$: Replace by $k$ the key of item $x$ in heap $H$.  The current key of $x$ must be greater than $k$.

\end{itemize}

The $\DecreaseKey$ operation requires knowing the location of the item $x$ whose key decreases in the heap that contains it.  If the heap implementation is endogenous (the items themselves are the heap nodes), then this location is the item itself.  If the heap implementation is exogenous (the heap nodes store the items), then the heap implementation must maintain the location of each heap item in its current heap.  If both $\DecreaseKey$ and $\Meld$ are supported operations, it is up to the user of the heap implementation to keep track of which heap contains which items.  This is an instance of the disjoint set union, or union-find problem~\cite{journals/jacm/Tarjan75}.  

We implement Dijkstra's algorithm using a heap containing the labeled vertices, each with a key equal to its current distance.  Each scanning step begins by doing a \DeleteMin{} on the heap to determine $v$, the next vertex to scan.  During the step, for each arc $vw$ such that $w$ is unlabeled it does a heap \Insert; for each arc $vw$ such that $w$ is labeled and $d(v)+c(vw) < d(w)$ it does a heap \DecreaseKey.  There is only one heap; the algorithm does not do any melds.  There is one heap \Insert{} and one heap \DeleteMin{} per vertex, and at most one heap \DecreaseKey{} per arc.  The algorithm runs in $\OO(m)$ time plus the time for the heap operations.  (Recall from~\cref{sec:graph-concepts} that we assume $n >2$, so $n=\OO(m)$.)  

\begin{lemma}\label{lem:forward-arc-comparisons}
During a run of Dijkstra's algorithm there are at most $F-n+1$ comparisons outside of heap operations and at most $F-n+1$ heap \DecreaseKey{} operations, where $F$ as defined in~\cref{sec:graph-concepts} is the maximum number of forward arcs for any distance order.
\end{lemma}
\begin{proof} Let $\{v_1=s, v_2,\dots,v_n\}$ be the vertices in the order they are scanned. By~\cref{lem:Dijkstra-scanning-order}, $L$ is a distance order.  During the scan of a vertex $v$, there is one comparison (outside of heap operations) for each arc $vw$ such that $w$ is labeled.  Each such arc $vw$ is a forward arc of $L$.  Depending on the outcome of this comparison, one heap \DecreaseKey{} operation may occur.  In addition, there are exactly $n-1$ forward arcs of $L$ that cause an insertion into the heap but no comparison outside of the heap insertion and no \DecreaseKey{} operation.  The lemma follows. 
\end{proof}

\begin{remark}
In the case of an undirected graph, the bound in~\cref{lem:forward-arc-comparisons} becomes $m-n+1$.
\end{remark}

Dijkstra's original implementation of his algorithm represented the heap by an array indexed by vertex, with the key of a vertex stored in the corresponding array position.  With this representation, each \Insert{} and each \DecreaseKey{} takes $\OO(1)$ time and comparisons, but each \DeleteMin{} takes $\Theta(n)$ time and comparisons, resulting in a tight bound of $\Theta(n^2)$ time and comparisons, which is optimal for dense graphs but far from optimal for sparse graphs.  Currently, the standard heap implementation for Dijkstra'a algorithm in practice is an implicit $d$-heap~\cite{Tarjan1983}.  This is a complete $d$-ary tree that is min-heap-ordered by key and stored in an array.  A commonly accepted good choice of $d$ is four.  With this implementation, or more generally with any heap implementation that supports each operation on an $n$-item heap in $\OO(\log n)$ time and comparisons, the worst case running time of Dijkstra's algorithm is $\OO(m\log n)$, as is the number of comparisons.

Fibonacci heaps \cite{fredman1987fibonacci} were invented to speed up Dijkstra's algorithm and other algorithms.  They support each heap operation except \DeleteMin{} in $\OO(1)$ amortized time and comparisons, and \DeleteMin{} on an $n$-item heap in $\OO(\log n)$ amortized time and comparisons.  This gives an $\OO(m+n\log n)$ worst-case time and comparison bound for Dijkstra's algorithm.  For all but very sparse graphs, this is an asymptotic improvement over the bound obtained with an implicit heap.  Furthermore, there are graphs for which this bound is best possible, not only for any implementation of Dijkstra's algorithm but for \emph{any} algorithm that finds a true distance order.  This follows from the lower bounds we develop in the next section.

For an implementation of Dijkstra's algorithm to achieve the stronger property of universal optimality, the heap it uses needs to have a bound for \DeleteMin{} that depends in a more fine-grained way on the sequence of heap operations.  This bound is the working-set bound, defined as follows.  If an item is deleted from a heap and later re-inserted, we treat it as a new item, so each item is only inserted and deleted once.  The \emph{working set} of an item $x$ in a heap is the set of items inserted into the heap from the time $x$ is inserted until the time $x$ is deleted.  The working set includes $x$ itself.  The \emph{working-set size} $W(x)$ of $x$ is the number of items in its working set.  A heap has the \emph{working-set bound} if each supported heap operation other than \DeleteMin{} takes $\OO(1)$ amortized time and comparisons, and each  \DeleteMin{} that returns an item $x$ takes $\OO(\log W(x))$ amortized time and comparisons.

In our efficiency analysis we shall assume that Dijkstra's algorithm is implemented with a heap that supports \MakeHeap{}, \FindMin{}, \Insert{}, \DecreaseKey{}, and \DeleteMin{} and has the working-set bound.  We do not require the \Meld{} operation.  In~\cref{sec:heap} we shall develop a heap that has the required efficiency.

\section{Complexity Models and Universal Optimality}\label{sec:complexity-models}

To study the complexity of the distance order problem we need a complexity model.  We shall consider two, the \emph{comparison model} and the \emph{time model}.  In the comparison model, the algorithm knows the arcs and vertices of the graph and has oracle access to the arc lengths.  It pays one for each comparison of two linear functions of arc lengths.  No other operations on arc lengths are allowed.  We say an algorithm for the distance order problem is \emph{universally optimal in comparisons} if, on any graph, it does a number of comparisons within a constant factor of the maximum number required, where the maximum is taken over every choice of arc lengths.

In the time model, the problem graph is represented by its vertex set, the source vertex $s$, and an incidence list of outgoing arcs for each vertex.  Each arc stores its length.  The algorithm must discover the arcs and their costs by traversing the incidence lists.
Given a vertex, the algorithm can in one unit of time access the first arc on an incidence list.  The access returns $\Null$ if the list is empty.  Given an arc $vw$, the algorithm can in one unit of time access the arc after $vw$ on the incidence list of $v$.  The access returns $\Null$ if there is no arc after $vw$ on the list.  As in the query model, the only operations allowed on arc lengths are comparisons of linear functions of arc lengths, but all arc lengths in such a function must be of previously accessed arcs.  A comparison takes one unit of time. (This model is unrealistically strong, since an actual algorithm must compute the linear functions of arc lengths that it is comparing, but this only makes our lower bounds stronger.)  We say an algorithm for the distance order problem is \emph{universally optimal in time} if, on any graph, it takes time within a constant factor of the maximum time required on the given graph, where the maximum is taken over every choice of arc lengths and every representation of the graph (every permutation of the incidence lists).  
  
The comparison model is strictly stronger than the time model, but less realistic.  An extreme example of the difference is a graph consisting of a single path $[s=v_1, v_2,\dots,v_n]$ and any number of arcs of the form $v_jv_i$ with $j>i$.  On such a graph the distance order problem takes zero comparisons in the query model but $\Omega(m)$ time in the time model.

In the next section we prove that the distance order problem on any graph takes $\Omega(m+\log D)$ time in the time model and $\Omega(F-n+1+\log D)$ comparisons in the comparison model, where $D$ is the number of distance orders of the problem graph and $F$ is the maximum number of forward arcs of any distance order.  In the three sections following the lower bound section, we prove that Dijkstra's algorithm implemented using an appropriate heap runs in $\OO(m+\log D)$ time on any graph, and hence is universally optimal in time; and that two augmented versions of the algorithm run in $\OO(m+\log D)$ time and do $\O(F-n+1+\log D)$ comparisons on any graph, and hence are universally optimal in both time and comparisons.

\section{Lower Bounds for Distance Order}\label{sec:lower-bounds}

For the distance order problem we prove one lower bound on time and two on comparisons, and then combine them.  We begin with a lower bound on time.

\subsection{A lower bound on time}

\begin{lemma}\label{lem:time-lower-bound}
In the time model, a correct deterministic algorithm for the distance order problem requires $\Omega(m)$ time on any graph.
\end{lemma}
\begin{proof}
Let $G$ be any graph and $L=[v_1,v_2,\dots,v_n]$ any distance order of $G$.  By~\cref{lem:tree-top-dist}, each vertex $w\neq s$ has an incoming forward arc.  We shall choose arc lengths for $G$ such that $L$ is the unique true distance order.  We shall also choose an order of the incidence lists of $G$.  We shall prove that if an algorithm for the distance order problem does not spend at least $\max\{n-2, m-2n+2\}$ units of time to solve the problem on $G$, then there is a closely related graph with the same vertex set as $G$ but such that $L$ is not a true distance order.  Furthermore on this graph the algorithm behaves the same as on $G$, and hence outputs $L$. Thus the algorithm is incorrect.

We choose $c(v_iv_j)=\max\{0,j-i\}$.  Then $d^*(v_i)=i-1$, so $L$ is the unique true distance order of $G$.  For each vertex $v_i$, if there is an arc $v_iv_j$ such that $j>i+1$, we move one such arc to the end of the incidence list of $v_i$.

Consider a run of a distance-order algorithm on $G$.  Suppose for some $v_i$ with $i < n-1$ the algorithm does not access the incidence list of $v_i$ at all.  If $v_i$ has an outgoing arc $v_iv_j$ with $j > i+1$, decrease the length of this arc to $1/2$.  If $v_i$ does not have such an outgoing arc, add such an arc, with length $1/2$.  Then $L$ is not a true distance order of the modified graph, since there is a path from $s=v_1$ to $v_j$ of length $i-1+1/2$ but every path from $s$ to $v_{i+1}$ has length at least $i$, so $v_j$ precedes $v_{i+1}$ in any correct distance order.  But the algorithm will output $L$ given the specified representation of the modified graph as input.  Thus the algorithm is incorrect.  We conclude that the time a correct algorithm takes, given the specified representation of $G$ as input, is at least $n-2$, since it must access at least $n-2$ incidence lists.

Suppose there is some vertex $v_i$ with $i<n-1$ such that the algorithm does not access \emph{all} arcs on the incidence list of $v_i$.  If the last arc on the incidence list of $v_i$, say $v_iv_j$, has $j > i+1$, decrease the length of this arc to $1/2$.  If there is no such arc on the incidence list of $v_i$, add such an arc with length $1/2$ as the last arc on the list.  List $L$ is not a true distance order of the modified graph, but the algorithm will output $L$ when given the representation of the modified graph as input.  Thus the algorithm is  incorrect.  We conclude that the time a correct algorithm takes, given the specified representation of $G$ as input, is at least $m-2n+2$.

The lemma follows, since our assumption $n>2$ implies $\max\{n-2, m-2n+2\}=\Omega(m)$.
\end{proof}

In the case of an undirected graph,~\cref{lem:time-lower-bound} holds, with a slightly different proof.  Each edge connecting two vertices $v_i$ and $v_j$ is represented by two arcs, $v_iv_j$ and $v_jv_i$, which are on the arc lists of $v_i$ and $v_j$, respectively.  For each arc $v_iv_j$ with $j > i+1$, a correct algorithm must access either $v_iv_j$ or $v_jv_i$, or one can get a contradiction as in the proof of~\cref{lem:time-lower-bound}.  This gives a lower bound of $m-n+1$ time, where $m$ is the number of edges in the original undirected graph.  If we pair the vertices so that each pair $v_i, v_j$ has $j> i+1$, then a correct algorithm must access the incidence list of $v_i$ or that of $v_j$, or there could be a pair of arcs $v_iv_j$ and $v_jv_i$, again allowing a contradiction as in the proof of~\cref{lem:time-lower-bound}.  Thus we obtain a lower bound of $\max\{\lfloor n/2 \rfloor,m-n+1\}$ time.

The lower bound on time for undirected graphs holds for \emph{every} representation of the given graph, not just for an adversarially chosen one.  A natural question is whether this is true for directed graphs as well.  The answer is yes if the graph can have multiple arcs but no if not.  If multiple arcs are allowed, the algorithm must access the entire incidence list of each vertex except the last one in true distance order, since otherwise there could be an unexamined forward arc that could contradict the correctness of the produced order.  If multiple arcs are not allowed, the situation is more complicated.  Let $v_1, v_2,\dots, v_n$ be the distance order produced by the algorithm.  For each $v_i$, a correct algorithm must either access the entire incidence list of $v_i$, or there must be an arc $v_iv_j$ for each $j>i$ and the algorithm must access each such arc: If the algorithm does not access all arcs out of $v_i$ and does not access an arc $v_iv_j$ for each $j>i$, there could be such an arc, which could contradict the correctness of the produced order.  The version of Dijkstra's algorithm presented in~\cref{sec:dijkstra} accesses \emph{all} arcs, but it is easy to modify it to match this tighter lower bound: When a vertex $v$ is scanned, the algorithm maintains a count of arcs out of $v$ that lead to unscanned vertices, and stops the scan when this count equals the total number of unscanned vertices.

An alternative version of the time model is to assume a more-primitive model of the input, in which the graph is given as a list of arcs together with their costs.  If the input is in this form, any correct algorithm must access $\min\{m, n(n-1)/2\}$ arcs to make sure it hasn't missed a forward arc.

\subsection{Lower bounds on comparisons}

Now we turn to comparisons.  We obtain a lower bound of $\Omega(\log D)$ comparisons by applying the standard information-theory lower bound argument for sorting by binary decisions.

\begin{lemma}\label{lem:D-lower-bound}
In the comparison model, a correct deterministic algorithm for the distance order problem does at least $\lceil\log D\rceil$ comparisons\footnote{Throughout this paper $\log$ without a base denotes the base-two logarithm.} on any graph in the worst case, and at least $\lfloor\log D\rfloor$ comparisons on any graph in expectation, if the arc weights are chosen according to a worst-case distribution.  
\end{lemma}
\begin{proof}
Without loss of generality, we can assume that each comparison tests whether a given linear function of arc lengths is less than, greater than, or equal to $0$.   We can model an algorithm for the distance order problem by a decision tree in which each node represents a comparison.  Each such node has three outcomes, depending on the result of the comparison.  For each possible distance order of $G$, consider a set of strictly positive arc lengths such that the given order is the unique true distance order and all true distances are distinct, such as the arc lengths defined in the proof of~\cref{lem:tree-top-dist}.  For each such set of arc lengths that results in an ``equal to $0$" outcome for some comparison in the decision tree, perturb the value of some arc length whose coefficient in the linear function is non-zero so that the outcome of the comparison changes to ``greater than $0$" or ``less than $0$".  Make the perturbation small enough so that the arc length remains strictly positive, no other non-zero comparison outcome for this set of arc lengths changes, and the true distances remain distinct and in the same order.   This is always possible, because arc lengths can be arbitrary non-negative real numbers, so we can make arbitrarily small changes.  (In fact, we can make all arc lengths rational.)

Now there are $D$ different sets of arc lengths, each of which must follow a distinct path in the decision tree and none of which has an ``equal to $0$" outcome.  The set of nodes in the decision tree that are on at least one such path form a binary tree that must contain at least $D$ leaves and hence must have a leaf of depth at least $\lceil\log D\rceil$, so the worst-case number of comparisons is at least $\lceil\log D\rceil$. Furthermore, if a leaf is chosen uniformly at random, its expected depth is at least $\lfloor\log D\rfloor$
, so if the set of arc weights is chosen uniformly at random from among the perturbed sets, the expected number of comparisons is at least $\lfloor\log D\rfloor$.

\end{proof}

The proof of~\cref{lem:D-lower-bound} yields a lower bound of $\lfloor\log_3 D\rfloor$ in the average case without perturbation of the arc lengths.  By Yao's principle~\cite{Yao77}, this lower bound holds to within a constant factor even for randomized algorithms, both Las Vegas and Monte Carlo.


\begin{remark}
 In the case of an undirected graph, the proof of~\cref{lem:D-lower-bound} holds as stated.  The comparisons are of linear functions of the edge weights.   
\end{remark}

\begin{lemma}\label{lem:F-lower-bound}
In the comparison model, a correct deterministic algorithm for the distance order problem does at least $F-n+1$ comparisons.
\end{lemma}
\begin{proof}
Let $L=[v_1,v_2,\dots,v_n]$ be a distance order of $G$ having $F$ forward arcs.  Give each forward arc $v_iv_j$ a length of $j-i$.  We shall specify the lengths of the non-forward arcs later: We shall choose all of these lengths to be the same, and to be sufficiently large that the behavior of the distance order algorithm depends only on the lengths of the forward arcs, as long as we do not change these lengths too much.  List $L$ is the unique true distance order and all true distances are distinct, for any choice of non-negative lengths of the non-forward arcs.

As in the proof of~\cref{lem:D-lower-bound}, consider the decision tree corresponding to the algorithm.  Assume that all the non-forward arcs have the same length, and that this length is very large.  Let $a\cdot c$ be the linear function compared to $0$ in some node of the decision tree, where $c$ is the vector of forward arc lengths and one variable representing the common value of the lengths of all non-forward arcs, and $c$ is a vector of real-valued coefficients.  If the coefficient of the non-forward-arc variable is $0$, we delete this term, resulting in a function of just the lengths of the forward arcs.  If the coefficient of the non-forward-arc variable is positive or negative, we replace the node in the decision tree by a direct branch to the ``greater than $0$" or ``less than $0$" outcome, respectively.  The result is a simplified decision tree with linear functions of variables for only the forward arcs.  Later we shall choose the length of all the non-forward arcs to be sufficiently large to justify this assumption.

Perturb the lengths of the forward arcs so that no comparison outcome on the path in the simplified decision tree corresponding to the specified arc lengths results in an ``equal to $0$" outcome, but such that all arc lengths remain strictly positive, $L$ remains the unique true distance order, and all true distances remain distinct.
This is always possible, because arc lengths can be arbitrary non-negative real numbers, so we can make arbitrarily small changes.  (In fact, we can make all arc lengths rational.)

Suppose that the path followed for the perturbed forward arc lengths in the simplified decision tree contains at most $F-n$ comparisons.  We shall construct a new set of forward arc lengths on which the algorithm does the same comparisons but for which $L$ is not a true distance order, making the algorithm incorrect.

We construct the new set of arc lengths as follows.  Let $a\cdot c>0$ or $a\cdot c < 0$ be one of the (at most $F-n$) comparison inequalities satisfied by the perturbed set of arc lengths on the path it takes in the simplified decision tree.  Here $c$ is the vector of forward arc lengths.  Convert this inequality into an equality by evaluating $a \cdot c$ for the given set of forward arc lengths and replacing the $0$ on the right-hand side with the resulting value.  Do this conversion for each of the inequalities on the path in the decision tree.  This produces a set of at most $F-n$ linear equations in the forward arc lengths.  For each $i$ such that $v_iv_{i+1}$ is an arc of $G$, add to the set of equations the equation $c(v_iv_{i+1})=\ell$, where $\ell$ is the length of $v_iv_{i+1}$ in the perturbed set of forward arc lengths.  After these at most $n-1$ additions, the result is a set of at most $F-1$ linear equations in the forward arc lengths.  This set of equations has at least one solution, the set of perturbed forward arc lengths.  Since it has at least one solution, and since there are fewer equations than variables, the set of solutions is a space with at least one dimension.  Choose a line in this space containing the known solution and move along this line in a direction that reduces the value of at least one variable.  Continue until some variable becomes $0$.

We now have a new set of forward arc lengths in which one or more equals $0$ and all are non-negative.  This new set of arc lengths follows the same path in the simplified decision tree as the original perturbed set of arc lengths, since each linear function on this path has the same value for both sets of arc lengths.  Furthermore no comparison has an ``equal to $0$" outcome.  We can slightly increase the values of all but one of the forward arcs of length $0$
in the new set of forward arc lengths without changing any of the comparison outcomes.  Furthermore we can do this in such a way that no two distances are the same.
Now we we have exactly one forward arc $v_iv_j$ with $j>i+1$ and length zero.  By making the common length of the non-forward arcs sufficiently large, we can guarantee that the original perturbed set of forward arc lengths and the new set of forward arc lengths follow the same path in the original decision tree, for the common large length of the non-forward arcs.  But $L$ is not a true distance order for the new set of arc lengths, since in a true distance order for the new set of arc lengths, $v_j$ must immediately follow $v_i$, or precede it.
Hence the distance order algorithm is incorrect. 
\end{proof}

In the case of an undirected graph, the proof of~\cref{lem:F-lower-bound} becomes simpler, because each comparison is of a linear function of the edge lengths with $0$: We do not need to treat forward and non-forward arcs separately.  Since for undirected graphs $F=m$, the lower bound is $m-n+1$ comparisons.

Since a lower bound on comparisons is also a lower bound on time,~\cref{lem:time-lower-bound} and \cref{lem:D-lower-bound} combine to give us our time lower bound, which holds for both directed and undirected graphs:

\begin{theorem}\label{thm:time-lower-bound}
In the time model, any correct deterministic algorithm for the distance order problem takes $\Omega(m+\log D)$ time.
\end{theorem}
\begin{proof}
Combining~\cref{lem:time-lower-bound} and~\cref{lem:D-lower-bound} gives us a lower bound of $\Omega(\max\{n-2,m-2(n-1),\log D\})=\Omega(m+\log D)$ for a directed graph, $\Omega(\max\{\lfloor n/2 \rfloor, m -n +1, \log D\}) = \Omega(m+\log D)$ for an undirected graph, since we are assuming $n>2$.
\end{proof}

Our two lower bounds on comparisons combine to give us our comparison lower bound:

\begin{theorem}\label{thm:comparison-lower-bound}
In the comparison model, any correct deterministic algorithm for the distance order problem takes $\Omega(F-n+1+\log D)$ comparisons on a directed graph, $\Omega(m-n+1+\log D)$ comparisons on an undirected graph.
\end{theorem}

\section{Efficiency of Dijkstra's Algorithm}\label{sec:dijkstra-efficiency}

In this section we analyze the worst-case running time and number of comparisons of Dijkstra's algorithm when implemented with a heap having the working-set bound.  We obtain a worst-case time bound of $\OO(m+\log D)$ and a worst-case comparison bound of $\OO(F+\log D)$, where $F$ as defined in~\cref{sec:graph-concepts} is the maximum number of forward arcs of a distance order of $G$.  The former matches the time lower bound of~\cref{thm:time-lower-bound}.  The latter matches the comparison lower bound of~\cref{thm:comparison-lower-bound} to within an additive term in $n$.  In the next two section we augment Dijkstra's algorithm in two different ways to produce two algorithms that match both lower bounds.

Given that the heap has the working-set bound, each heap insertion and each \DecreaseKey{} operation takes $\OO(1)$ amortized time and comparisons.  It follows from the discussion in~\cref{sec:dijkstra-heap} and from~\cref{lem:forward-arc-comparisons} that Dijkstra's algorithm takes $\OO(m)$ time plus the time for the \DeleteMin{} operations, and does $\OO(F)$ comparisons plus the comparisons done by the \DeleteMin{} operations.  Thus all we need to bound is the time taken and the number of comparisons done by the \DeleteMin{} operations.  The time taken by a \DeleteMin{} is bounded by $\OO(1)$ plus a constant times the number of comparisons it does.  Thus we just need to bound the number of comparisons done by \DeleteMin{} operations. 

Our tool for doing this is Lemma 2.3 of~\citet{vanderhoog2024} that gives a lower bound on the number of topological orders of a certain class of graphs.  We restate their lemma in a form convenient for us.  We need some definitions.  For any integers $i$ and $j$, such that $i \leq j$, we denote by $[i, j]$ the interval of integers $\{i, i+1,\dots, j\}$.  Let $\{\,[a_i,b_i] \mid 1 \leq i \leq k\,\}$ be a set of $k$ intervals of integers, each a subset of $[1,k]$.  (It is not a coincidence that the number of intervals is the same as the size of the ground set.)  The DAG (directed acyclic graph) $I$ associated with this set of intervals is the graph whose vertices are the intervals $[a_i,b_i]$ and whose arcs are the pairs $[a_i,b_i], [a_j,b_j]$ such that $b_i < a_j$.   We denote by $D(I)$
the number of topological orders of $I$.  We restate Lemma 2.3 of~\citet{vanderhoog2024} in a form more convenient for our use.     

\begin{lemma}\label{lem:interval-bound} \textup{\cite{vanderhoog2024}}
Let $I$ be the DAG associated with the set of intervals $\{\,[a_i, b_i] \mid 1 \leq i \leq k\,\}$, each a subset of $[1, k]$.  Then $\sum_{i=1}^k \log (b_i-a_i+1) = \OO(\log D(I))$.
\end{lemma}

\begin{remark}
In the original lemma, the intervals are open intervals that are subsets of the interval of real numbers $[0,k]$, each interval having a length of at least $1$.  The mapping $[a_i,b_i] \rightarrow (a_i-1, b_i)$ converts our restatement into an instance of the original lemma. 
\end{remark}

We use~\cref{lem:tree-top-dist} and~\cref{lem:interval-bound} to bound the sum of the logarithms of the working-set sizes in a run of Dijkstra's algorithm.

\begin{lemma}\label{lem:working-set-log-T}
Given a run of Dijkstra's algorithm on a graph $G$, let $W(v)$ for $v \in V$ be the working-set size of $v$; that is, the number of vertices inserted into the heap from the time $v$ is inserted until $v$ is deleted, including $v$. Then $\sum_{v\in V}\log W(v)= \OO(\log D)$. 
\end{lemma}

\begin{proof}
Consider a run of Dijkstra's algorithm.  Let $[v_1, v_2,\dots, v_n]$ be the sequence of vertices in the order they are inserted into the heap.  For each vertex $v_i$, let $[a_i=i, b_i]$ be the set of indices of vertices inserted into the heap from the time $v_i$ is inserted until the time $v_i$ is deleted.  Then $W(v_i)= b_i-a_i+1$.  Let $I$ be the DAG associated with the set of intervals $\{\,[a_i, b_i] \mid 1 \leq i \leq n\,\}$.  For each $i$ from $2$ to $n$, let $u_iv_i$ be the arc whose processing caused $v_i$ to become labeled.  Then the set of arcs $\{\,u_iv_i \mid 2\le i\le n \}$ forms a spanning tree $T$ of $G$ rooted at $s$, specifically the search tree generated by the run of Dijkstra.  We claim that each topological order of $I$ gives a distinct topological order of $T$ if we replace each interval $[a_i,b_i]$ in the order by $v_i$.

To prove the claim, we observe that if $v_iv_j$ is an arc of $T$, then $v_i$ is deleted from the heap before $v_j$ is inserted.  Hence $b_i < a_j$, so there is an arc in $I$ from the interval $[a_i,b_i]$ to the interval $[a_j,b_j]$.  It follows that if an order of intervals in $I$ is topological, so is the corresponding order of vertices of $T$.

By~\cref{lem:tree-top-dist}, each topological order of $T$ is a distance order of $G$.  By the claim, the number of distance orders of $G$ is at least the number of topological orders of $I$.  The lemma follows from~\cref{lem:interval-bound}.      
\end{proof}

By~\cref{lem:working-set-log-T}, the number of comparisons done by the \DeleteMin{} operations during a run of Dijkstra's algorithm is $\OO(n+\log D)$.  Combining our bounds we obtain our first main result:

\begin{theorem}\label{thm:Dijkstra-bounds}
Dijkstra's algorithm implemented with a heap having the working-set bound runs in $\OO(m+\log D)$ time and does $\OO(F+\log D)$ comparisons on a directed graph, $\OO(m+\log D)$ comparisons on an undirected graph.  Hence it is universally optimal in time, and universally optimal in comparisons up to an additive term in $n$.
\end{theorem}

\section{Dijkstra's Algorithm with Lookahead}\label{sec:dijkstra-lookahead}

By~\cref{thm:Dijkstra-bounds}, Dijkstra's algorithm implemented with a heap having the working-set bound does $\OO(F+\log D)$ comparisons on a directed graph, but by~\cref{thm:comparison-lower-bound} the number of comparisons required to solve the distance-order problem is $\Omega(F-n+1+\log D)$.  The gap between the upper and lower bounds is $\Theta(n)$, which is significant if $\log D$ is small.  Consider in particular the example in~\cref{fig:broom}.  In this section we explore when $\log D$ can be small compared to $n$.  Then we augment Dijkstra's algorithm to reduce the bound on comparisons to $\Omega(F-n+1+\log D)$, making it tight.

\subsection{Bottlenecks}\label{sec:bottlenecks}

Recall from~\cref{sec:graph-concepts} the following definitions: the \emph{level} $\el(v)$ of a vertex $v$ is the minimum number of vertices on a path from $s$ to $v$, and a \emph{bottleneck} is a vertex that is the only one on its level.

\begin{lemma}\label{lem:many-bottlenecks}
If $G$ has $b$ bottlenecks, then $\log D \geq (n-b)/2$.
\end{lemma}
\begin{proof}
Let $V_i$ be the set of vertices on level $i$, and let $\ell$ be the number of levels.  Since every level that does not contain a bottleneck contains at least two vertices, $2(\ell-b) \leq n-b$.  This gives $\ell \leq (n+b)/2$.  Let $T$ be a breadth-first spanning tree of $G$.  Then if $vw$ is an arc of $T$, $\el(w) = \el(v)+1$.  The number of topological orders of $T$ is at least $\prod_{i=1}^{\el}|V_i|!$, since one can form such an order by ordering the vertices in increasing order by level, and within each level ordering the vertices arbitrarily.  By~\cref{lem:tree-top-dist}, each topological order of $T$ is a distance order, so  \[D \geq \prod_{i=1}^{\el}|V_i|! \geq \prod_{i=1}^{\el}2^{|V_i| - 1}= 2^{n-\el}.\]  Taking logarithms of both sides of the inequality and combining the result with the inequality $\ell \leq (n+b)/2$ gives the lemma.
\end{proof}

\cref{lem:many-bottlenecks} implies that Dijkstra's algorithm is universally optimal in comparisons unless the number of bottlenecks in the graph is at least $(1-\epsilon)n$, where $\epsilon$ is any positive constant.  To handle graphs with such a large number of bottlenecks, we shall exploit the following two lemmas about such vertices.  Recall the following definitions from~\cref{sec:graph-concepts}: A vertex $v$ \emph{dominates} a vertex $w\neq v$ if $v$ is on every path from $s$ to $w$; an arc $wv$ is \emph{useless} if $v$ dominates $w$. 

\begin{lemma}\label{lem:bottleneck-useless}
Let $v$ be a bottleneck.  Then $v$ dominates all vertices on higher levels, and if $wv$ is an arc such that $\ell(v) < \ell(w)$, then $wv$ is useless.
\end{lemma}
\begin{proof}
If $xy$ is an arc, $\ell(y) \leq \ell(x)+1$.  Hence any path from $s$ to a vertex $x$ must contain at least one vertex on each level from $\el(s)=1$ to $\el(x)$.  Thus if $v$ is a bottleneck and $\el(v)<\el(w)$, any path from $s$ to $w$ contains $v$.  This gives both parts of the lemma. 
\end{proof}

We define a bottleneck to be \emph{marked} if the next-higher level contains at least two vertices, and \emph{unmarked} otherwise.  Equivalently, an unmarked bottleneck is one that is either on the highest level or has a bottleneck on the next-higher level.

\begin{lemma}\label{lem:bottleneck-dist}
If $v$ is an unmarked bottleneck that is not on the highest level, there is an arc $vw$ such that $d^*(w)=d^*(v)+c(vw)$.
\end{lemma}
\begin{proof}
Let $w$ be the unique vertex on level $\ell(v)+1$.  Since $v$ dominates $w$ and $w$ has level exactly one higher than $v$, any simple path from $s$ to $w$, and in particular a shortest path, contains an arc $vw$.  For this arc, $d^*(w)=d^*(v)+c(vw)$.   
\end{proof}

Suppose there are $k$ marked bottlenecks.  Then the sequence of bottlenecks in increasing order by level consists of $k+1$ subsequences, each subsequence except the last one ending with a marked bottleneck.  The bottlenecks in each subsequence are on consecutive levels.  By~\cref{lem:bottleneck-dist}, once we know the true distance of the first bottleneck in a subsequence, we can compute the true distance of all the other bottlenecks in the subsequence by scanning the bottlenecks of the subsequence in increasing order by level.  We use this idea in the next section to develop a variant of Dijkstra's algorithm that inserts only the non-bottlenecks into the heap: It handles the bottlenecks separately.  In general, the algorithm scans the vertices in a different order than Dijkstra's algorithm does: It can scan bottlenecks before scanning non-bottlenecks having smaller true distance.  We call the resulting algorithm \emph{Dijkstra with lookahead}.

\subsection{Dijkstra with lookahead}

Dijkstra with lookahead begins by finding all the bottlenecks and determining the marked ones.  To find the bottlenecks, it computes the level of every vertex by doing a breadth-first search from $s$ and forms a list of the bottlenecks in increasing order by level. It marks each bottleneck $v$ such that $\ell(v)+1$ contains at least two vertices.  Finding and marking the bottlenecks takes $\OO(m+n)$ time.

Having found and marked the bottlenecks, the algorithm runs Dijkstra's algorithm using a heap with the working-set bound, but without inserting the bottlenecks into the heap.  Instead, it maintains an array $B$ of bottlenecks that have not yet been added to the list $L$ of vertices sorted by distance.  It initializes $B$ to contain the bottlenecks in increasing order by level, up to and including the first marked one, or all the bottlenecks if none are marked.  When $B$ becomes empty (because all its vertices have been added to $L$), the algorithm refills $B$ by adding to it the bottlenecks not in $L$ in increasing order by level, up to and including the first marked one, or all the bottlenecks not on $L$ if none are marked.  Once all the bottlenecks have been added to $L$, the algorithm behaves exactly like Dijkstra's algorithm.

The algorithm builds a shortest path tree, by computing the parent $p(w)$ in the tree of each vertex $w\neq s$.  What the algorithm does at each step depends on whether $B$ or the heap $H$ is empty, and if both are non-empty on whether the smallest-distance vertex not yet on $L$ is in $B$ or in $H$. The algorithm uses the following version of vertex scanning:

\begin{itemize}
    \item[] $\Scan(v)$:  Mark $v$ scanned. Process each arc $vw$ by applying the appropriate one (if any) of the following two cases to it:
    \subitem[i] $w$ is unlabeled.  Mark $w$ labeled. Set $d(w) \gets d(v)+ c(vw)$, set $p(w)\gets v$, and if $w$ is not a bottleneck, insert $w$ into $H$ with key $d(w)$;
    \subitem[ii] $w$ is labeled and $d(v)+c(vw) < d(w)$. Set $d(w) \gets d(v)+ c(vw)$, set $p(w)\gets v$, and if $w$ is not a bottleneck, decrease to $d(w)$ the key of $w$ in $H$. 
\end{itemize}

The complete algorithm does the following: Construct a list of the bottlenecks in increasing order by level and mark those such that the next-higher level contains at least two vertices.  Initialize the state of each vertex to unlabeled.  Initialize list $L$ and heap $H$ to empty.  Initialize array $B$ to contain the bottlenecks in increasing order by level up to and including the first marked one, or all of them if none are marked.  Initialize $d(s)\gets 0$ and $d(v)\gets \infty$ for $v \neq s$.  Repeat the applicable one of the following two cases until $B$ and $H$ are empty:

\begin{itemize}

\item [] Case 1: $H$ is non-empty, and either $B$ is empty or its minimum-level vertex has current distance greater than that of $\FindMin(H)$.  Set $v \gets \DeleteMin(H)$, scan $v$, and add $v$ to $L$.

\item [] Case 2: Case 1 does not apply.  If the lowest-level vertex on $B$ is not yet scanned, scan each vertex on $B$ in increasing order by level.  Now all vertices in $B$ have their current distance equal to their true distance.  Apply the appropriate one of the following subcases:

\subitem Subcase 2a:  $B$ is non-empty, and either $H$ is empty or the maximum-level vertex in $B$ has current distance at most that of $\FindMin(x)$.  Move the bottlenecks in $B$ to $L$ in increasing order by level.  Once $B$ is empty, refill $B$ with the not-yet-scanned bottlenecks in increasing order by level, up to and including the first such bottleneck that is marked.  If all not-yet-scanned bottlenecks are unmarked, add all of them to $B$; if all bottlenecks have been scanned, $B$ remains empty and stays empty for the duration of the algorithm.

\subitem Subcase 2b: Subcase 2a does not apply.  Set $v \gets \FindMin(H)$.  Find the vertex $x$ of largest $d(x)$ in $B$ such that $d(x) \leq d(v)$.  To find $x$, do an exponential/binary search in $B$ starting at $p(v)$ if $p(v)$ is in $B$ or at the first vertex in $B$ if not. Compare $d(v)$ to the current distance of the first, second, fourth, eighth, \dots{} vertex in $B$ after the starting vertex until finding two consecutively compared vertices $y$ and $z$ in $B$ such that $d(y) \leq d(v) < d(z)$.  Then do a binary search on the set of vertices in $B$ between $y$ and $z$ to find $x$.  If $x$ is the $j$-th vertex in $B$ after the starting position, this search takes $\OO(1+\log j)$ time and $\OO(1+\log j)$ comparisons.  Move the bottlenecks in $B$ up to and including $x$ to $L$, in increasing order by level.
\end{itemize}

Dijkstra with lookahead is correct and efficient on any graph, as we shall show. By~\cref{lem:many-bottlenecks}, the added complication of handling the bottlenecks separately is unnecessary unless most of the vertices are bottlenecks.

\subsection{Correctness of Dijkstra with lookahead}

\begin{theorem}\label{thm:Dijkstra-look-correct}
Dijkstra with lookahead is correct.
\end{theorem}
\begin{proof} We begin the proof with some general observations about the behavior of the algorithm.  The algorithm differs from Dijkstra's algorithm only in that it can scan bottlenecks early.  As in Dijkstra's algorithm, $d(v)$ for each $v$ is the length of some path from $s$ to $v$ and never increases.  An induction on the number of iterations shows that each vertex will eventually be scanned, and later added to $L$.  At the beginning of an iteration, $B$ is non-empty unless all bottlenecks have been scanned and moved to $L$, in which case $B$ remains empty and only Case 1 applies for the duration of the algorithm.  When a set of bottlenecks is added to $B$ during the initialization or in Case 2a, all the bottlenecks in the set are unscanned.  During the next iteration of Case 2, they all become scanned.  Let $v$ be the unscanned bottleneck of lowest level.  Since $v$ dominates all vertices on higher levels, until $v$ is scanned every vertex on a higher level is unlabeled.

We prove by induction on the number of iterations of Cases 1 and 2 that the algorithm maintains the following invariant: When a vertex is scanned, its current distance is its true distance; all vertices on $L$ have current distance at most that of all vertices not on $L$; and vertices are added to $L$ in non-decreasing order by true distance. 

The invariant holds before any iterations.  Suppose the invariant holds before an iteration of Case 1.  Then the vertex $v$ deleted from $H$ has smallest current distance among all vertices not on $L$, and its current distance is no smaller than that of any vertex on $L$.  If $d(v) > d^*(v)$, there would have to be a pair of vertices $x$ and $y$ with $x$ on $L$ but $y$ not on $L$ and such that $d^*(x)+c(xy)<d(y)$.  But this would contradict the induction hypothesis, which states that $d(x)=d^*(x)$ and $x$ is scanned.  Hence $d(v)=d^*(v)$.  Deleting $v$ from $H$ and moving it to $L$ preserves the induction hypothesis.

Suppose the invariant holds before an iteration of Case 2.  Let $v$ be the lowest-level vertex on $B$.  By the argument for Case 1, $d(v)=d^*(v)$.  If $v$ is unscanned at the beginning of Case 2, each successive scan of a vertex on $B$ sets the current distance of the next vertex on $B$ equal to its true distance by~\cref{lem:bottleneck-dist}.  Hence when each such bottleneck is scanned, its current distance equals its true distance.  Thus these scans preserve the invariant.

Whether Subcase 2a or Subcase 2b applies, each successive bottleneck moved from $B$ to $L$ has smallest distance among the vertices not yet on $L$.  Hence each such move preserves the invariant.

To complete the proof, we need to show that $L$ is a distance order.  For each $v\neq s$, let $p(v)$ be the parent of $v$ in the shortest path tree computed by the algorithm.  Then $p(v)$ is added to $L$ before $v$.  This is immediate if $v$ is a bottleneck or $p(v)$ is not a bottleneck, because bottlenecks are added to $L$ in increasing order by level, and each non-bottleneck is added to $L$ immediately after it is scanned.  Suppose $p(v)$ but not $v$ is a bottleneck.  Since $d^*(p(v)) \leq d^*(v)$, and the tie-breaking rule for moving vertices to $L$ moves a bottleneck before it moves any non-bottleneck of equal distance, $p(v)$ will be added to $L$ before $v$.  This is true even if $v$ is in the heap before $p(v)$ is added to $B$, since in this case $v$ cannot be deleted from the heap until all bottlenecks up to and including $p(v)$ in increasing order by level have been added to $B$, had their true distances computed, and been moved to $L$.  By~\cref{lem:tree-top-dist}, $L$ is a distance order. 
\end{proof}

\subsection{Efficiency of Dijkstra with lookahead}

Now we study the running time and number of comparisons done by the algorithm.  In our analysis we shall assume that $D$, the number of distances orders of $G$, is at least $2$.  If $D=1$, $G$ consists of a path of bottlenecks, all unmarked, and possibly some useless arcs.  On such a graph, Dijkstra with lookahead will run in $\OO(m)$ time and do no comparisons, and hence is universally optimal in both time and comparisons.

We begin by bounding the time spent and the number of comparisons done outside of the heap operations and the searches of $B$ in Subcase 2b.  The time to find and mark the bottlenecks is $\OO(m)$.  Each vertex is added to $H$ or $B$ once, scanned once, and deleted from $H$ or $B$ and moved to $L$ once.  The total time for these computations is $\OO(m)$ plus the time for heap operations and searches of $B$.  Let $b$ be the number of bottlenecks.  There are $n-b$ non-bottlenecks and at most $n-b$ marked bottlenecks, since each marked bottleneck has a non-bottleneck on the next-higher level.
Each iteration of Case 1 does at most one comparison outside of heap operations, and it deletes one non-bottleneck from the heap.  Hence the iterations of Case 1 take $\OO(n-b)$ additional time and comparisons, not counting heap operations.  Each iteration of Subcase 2a does at most one comparison outside of heap operations and deletes a marked bottleneck from $B$, unless it deletes the last bottleneck from $B$, after which $B$ remains empty.  The total additional time and number of comparisons in iterations of Subcase 2a outside of heap operations is thus $\OO(n-b+1)$.  Each iteration of Subcase 2b is followed immediately by an iteration of Case 1. Hence there are at most $n-b$ iterations of Subcase 2b, which take a total of $\OO(n-b)$ additional time and comparisons, not counting heap operations and searches of $B$.

The algorithm does $n-b$ heap insertions, which is $\OO(\log D)$ by~\cref{lem:many-bottlenecks}.  The proof of~\cref{lem:forward-arc-comparisons} shows that Dijkstra with lookahead scans the vertices in a distance order.  It follows that Dijkstra with lookahead does at most $F-n+1$ heap \DecreaseKey{} operations, which take $\OO(F-n+1)$ time and comparisons, including each comparison done to determine whether to do a \DecreaseKey{}.

We conclude that the algorithm takes $\OO(m)$ time and does $\OO(F-n+1+\log D)$ comparisons, not including the time and comparisons for heap \DeleteMin{} operations and for searches of $B$ in Subcase 2b.  The next two lemmas bound the time taken and comparisons done by the \DeleteMin{} operations and the searches of $B$.

\begin{lemma}\label{lem:lookahead-deletions}
In a run of Dijkstra with lookahead, the time taken and number of comparisons done by the \DeleteMin{} operations is $\OO(\log D)$.
\end{lemma}
\begin{proof}
Consider a run of Dijkstra with lookahead.  For the purpose of the analysis, consider just the sequence of heap insertions, heap deletions, and vertex scans.  Precede each scan of a bottleneck $v$ by an insertion of $v$ into the heap followed by an immediate deletion of $v$ from the heap.  Such fictitious insertions and deletions only increase the working-set sizes of the non-bottlenecks.  Each bottleneck has a working-set size of $1$.  The proof of~\cref{lem:working-set-log-T} applies to this fictitious run to show that $\sum_{v\in V}\log W(v) = \OO(\log D)$, where $W(v)$ for $v \in V$ is the working-set size of $v$; that is, the number of vertices inserted into the heap from the time $v$ is inserted until $v$ is deleted, including $v$.  The lemma follows from this inequality and~\cref{lem:many-bottlenecks}. 
 \end{proof}

\begin{lemma}\label{lem:lookahead-search-time}
In a run of Dijkstra with lookahead, the searches of $B$ in Case 2b take $\OO(\log D)$ time and comparisons.  
\end{lemma}
\begin{proof}

Let $v$ be a non-bottleneck that is returned by the \FindMin{} during some iteration of Case 2b.  For each $v\neq s$, let $p(v)$ be the parent of $v$ in the shortest path tree computed by the algorithm.   Let $B(v)$ be the set of bottlenecks moved from $B$ to $L$ during this iteration of Case 2b, not including those of smaller level than $p(v)$ if $p(v)$ is moved from $B$ to $L$ during this iteration.  The exponential/binary search in this iteration of Case 2b takes $\OO(1+\log|B(v)|)$ time and comparisons.  Let $B(v)$ be the empty set for all other non-bottlenecks.  The sets $B(v)$ are disjoint, and if non-bottleneck $v$ is added to $L$ before non-bottleneck $w$ on $L$, then all vertices in $B(v)$ are added to $L$ before all vertices in $B(w)$.  Suppose $B(v)$ is non-empty.  Then $v$ must be in the heap before the iteration of Case 2b that moves the vertices in $B(v)$ to $L$.  Hence vertex $p(v)$ is either in $B(v)$ or was previously added to $L$, and hence is added to $L$ before $v$ is added to $L$.  Let $T$ be the shortest path tree built by the algorithm.  We can obtain a topological order of $T$, and hence by~\cref{lem:tree-top-dist} a distance order, by ordering all the bottlenecks in increasing order by level and inserting each non-bottleneck vertex $v$ into $B(v)$ in any of the $|B(v)|+1$ possible ways, including just before the first vertex in $B(v)$ and just after the last one.  (If $B(v)$ is empty, there is only one possible insertion position, just after the vertex added to $L$ just before $v$.)
This gives $D \geq \prod_v (|B(v)|+1)$.  Hence $\log D \geq \sum_v \log(|B(v)|+1)$. It follows from~\cref{lem:many-bottlenecks} that the time and number of comparisons spent on searches of $B$ is $\OO(\log D)$. 
\end{proof}

Combining our bounds gives us our second main result:

\begin{theorem}\label{thm:Dijkstra-lookahead-bounds}
Dijkstra's algorithm with lookahead implemented with a heap having the working-set bound runs in $\OO(m+\log D)$ time and does $\OO(F-n+1+\log D)$ comparisons on a directed graph, $\OO(m-n+1+\log D)$ comparisons on an undirected graph.  Hence it is universally optimal in both time and comparisons.
\end{theorem}

\begin{remark}
If the problem must be solved repeatedly for a fixed graph with different sets of arc lengths, the bottlenecks need only be found and marked once.
\end{remark}

\section{A recursive version of Dijkstra's algorithm}
\label{sec:dijkstra-recursive}

In this section we develop an alternative extension of Dijkstra's algorithm that is universally optimal in both time and comparisons.  The idea is to apply Dijkstra's algorithm recursively.
Recursive Dijkstra prioritizes the computation of the true distances of the bottleneck vertices.  The bottlenecks have a natural order by level; by~\cref{lem:bottleneck-useless}, each bottleneck dominates all vertices on higher levels, including the bottlenecks on higher levels.  We initiate a run of Dijkstra starting at $s$, which is the first bottleneck vertex.  We continue this run until the second bottleneck vertex, say $v$, is about to be scanned.  At this point we suspend the first run of Dijkstra and begin a second run, with $v$ as the start vertex and with its true distance, which is now known, as its initial current distance.  This run uses a new heap associated with $v$.  When the second run is about to scan the third bottleneck vertex, we suspend the second run and start a third run starting at this third bottleneck vertex.  We repeat until we have computed the true distance of every bottleneck vertex.

We then need to compute the true distances of all the non-bottlenecks, and to arrange all the vertices in non-decreasing order by distance.  We observe that any vertex scanned during any of the partial Dijkstra runs has its current distance equal to its true distance after the partial run, so we do not need to recompute these distances.  Furthermore, when scanning a vertex $x$ during the Dijkstra run that starts at some bottleneck vertex $w$, if an arc $xy$ leads to a vertex in the heap for some other bottleneck vertex $v$, it will be the case that $\ell(v)<\ell(w)$, and it is safe to decrease the key of $y$ in the heap associated with $v$ if the arc $xy$ provides a shorter path to $y$.  We do such decreases.

To compute the true distances of the remaining non-bottleneck vertices, we finish the run of Dijkstra for the highest-level bottleneck.  Then we resume the run for the next-highest-level bottleneck vertex and complete it.  We continue in this way, completing the Dijkstra runs of all the bottleneck vertices in decreasing order by level, until the run for the source $s$ has been completed.  Each higher-level run may do some \DecreaseKey{} operations on vertices in heaps for lower-level runs, but this is exactly what is needed to make the overall distance computation correct.

The final task is to construct a list $L$ of the vertices in non-decreasing order by distance.  To do this, we initialize the list $L$ to contain the bottlenecks in increasing order by level.  During the Dijkstra runs, when the processing of an arc $vw$ decreases the current distance of a non-bottleneck $w$, we set $p(w)\gets v$.  Thus when $w$ is scanned, $p(w)$ is the parent of $w$ in the shortest path tree defined by the Dijkstra runs.  If $w$ is not a bottleneck, just after it is scanned we insert $w$ into $L$ by doing a search in $L$ that starts from $p(w)$ and finds the first vertex $x$ after $p(w)$ with true distance at least that of $w$.  We insert $w$ into $L$ just before $x$.  If such an $x$ does not exist, we insert $w$ after the last vertex on $L$.  This guarantees that $L$ is a true distance order.

\subsection{Recursive Dijkstra}

Let us provide the full details of the recursive Dijkstra algorithm.  The algorithm uses the following version of vertex scanning, where vertex $u$ is the source vertex of the currently active Dijkstra run:

\begin{itemize}
    \item[] $\Scan(v)$: Mark $v$ scanned.  If $v$ is not a bottleneck, insert $v$ into $L$ just before the first vertex after $p(v)$ that has current distance at most $d(v)$; if there is no such vertex, insert $v$ after the last vertex on $L$. Process each arc $vw$ by applying the appropriate one (if any) of the following two cases to it:
    \subitem [i] if $w$ is unlabeled, mark $w$ labeled, set $d(w) \gets d(v)+ c(vw)$,  insert $w$ into $H(u)$ with key $d(w)$, and set $p(w)\gets v$;
    \subitem [ii] if $w$ is labeled and $d(v)+c(vw) < d(w)$, set  $d(w) \leftarrow d(v)+ c(vw)$, decrease to $d(w)$ the key of $w$ in the heap containing it, and set $p(w)\gets v$. 
\end{itemize}

We state recursive Dijkstra as an initialization followed by a call of a recursive method.  The initialization consists of finding all the bottleneck vertices, setting $d(s)\gets 0$ and $d(v)\gets \infty$ for $v \neq s$, initializing all vertices to be unlabeled, and initializing the distance-ordered list $L$ of vertices to contain the bottlenecks in increasing order by level.  After the initialization, the algorithm calls $\Dijkstra(s)$, where $\Dijkstra(u)$ is defined as follows:

\begin{itemize}
\item [] $\Dijkstra(u)$: Initialize a heap $H(u)$ to empty.  Do $\Scan(u)$.  While $H(u)$ is non-empty, do the following: Set $v\gets \DeleteMin(H(u))$.  If $v$ is a bottleneck then call $\Dijkstra(v)$; otherwise, do $\Scan(v)$. 
\end{itemize}

To support the local searching needed to make the insertions into $L$ efficient, we represent $L$ by a homogeneous finger search tree \cite{HuddlestonM82}.  Such a tree supports the insertion of a new item $k$ positions away from a given item (in symmetric order) in $\O(1+\log k)$ time.

Even though recursive Dijkstra uses several heaps rather than just one, there are no \Meld{} operations, so the efficient heap implementation we develop in~\cref{sec:heap} (which does not support \Meld) can be used in recursive Dijkstra. 

\subsection{Correctness of recursive Dijkstra}

\begin{theorem}
Recursive Dijkstra is correct.
\end{theorem}
\begin{proof}
We need to prove that recursive Dijkstra computes the true distances, and that it constructs a distance order.  To prove the former, let $v_1, v_2,\dots, v_b$ be the bottleneck vertices in increasing order by level.  By~\cref{lem:bottleneck-useless}, for $i < j$ there is no arc from a vertex inserted into $H(v_i)$ to a vertex inserted into $H(v_j)$.  Consider the first half of the computation, up to and including the scan of $v_b$.  A proof by induction on increasing $i$ shows that during the run of Dijkstra with source $v_i$, when a vertex is scanned, up to and including the scan of $v_{i+1}$, its current distance is its true distance, since this run does the same computations as standard Dijkstra if it is run with source $v_i$, except that $d(v_i)$ is initialized to the true distance of $v_i$ from $s$ instead of to $0$, so all the current distances for vertices added to $H(v_i)$ during the run are offset by $d(v_i)$.

Now consider the second half of the computation, after the scan of $v_b$.  We claim that when a vertex is scanned after the scan of $v_b$, its current distance is its true distance.  We prove the claim by induction on decreasing $i$.  The claim is true during completion of the run of Dijkstra with source $v_{b-1}$, since this entire run is just like standard Dijkstra but with $d(v_b)$ equal to the true distance from $s$ to $v_b$ rather than $0$.  Suppose the claim is true up to and including completion of the run of Dijkstra with source $v_{i+1}$.  Since all arc lengths are non-negative, all vertices scanned during the runs of Dijkstra with sources $v_{i+1}, v_{i+2},\dots, v_b$ have true distance at least $d(v_{i+1})$, and any update to the current distance of a vertex in $H(v_i)$ also has a value that is at least $d(v_{i+1})$.  Since $v_{i+1}$ was the last vertex removed from $H(v_i)$ before the run of Dijkstra with source $v_{i+1}$ was begun, all vertices in $H(v_i)$ when this happened had current distances at least $d(v_{i+1}$, and the updates of these current distances done by the runs with sources $v_{i+1}, v_{i+2},\dots, v_b$ preserve this.  It follows that during the entire run of Dijkstra with source $v_i$, vertices are deleted from $H(v_i)$ in non-decreasing order by current distance.

This implies that when a vertex is deleted from $H(v_i)$, its current distance is its true distance.  Suppose not.  Let $w$ be a vertex for which this is not true, and among such vertices one of minimum true distance, with a tie broken in favor of a vertex with fewest arcs on a shortest path from $s$.  Let $vw$ be the last arc on a shortest path of fewest arcs from $s$ to $v$.  Then $v=v_i$ or $v$ is dominated by $v_i$, so $v$ is scanned during the run of Dijkstra with source $v_j$ for some $j \geq i$.  If $j>i$, the current distance of $v$ is its current distance when it is scanned, and it is scanned before $w$, so when it is scanned the current distance of $w$ becomes equal to its true distance, a contradiction.  This is also true if $j=i$: By the choice of $w$, $v$ when scanned has its current distance equal to its true distance, and this is strictly less than the current distance of $w$ when it is scanned, so $v$ must be scanned before $w$.  The claim follows by induction.  That is, when a vertex is scanned, its current distance equals its true distance.

To finish the proof of correctness, we must show that the list $L$ constructed by the algorithm is a distance order.  If $v$ is a non-bottleneck, $v$ has true distance at least that of $p(v)$, its parent in the shortest path tree found by the algorithm.  It follows that each insertion into $L$ maintains $L$ in non-decreasing order by true distance.  We claim that each vertex $w\neq s$ occurs in $L$ after its parent in the shortest path tree $T$ defined by the Dijkstra runs.  The claim implies that $L$ is a topological order of $T$ and hence a distance order by~\cref{lem:tree-top-dist}.  The claim is true for non-bottlenecks by the way insertion into $L$ is done.   To prove the claim for bottlenecks, let $w$ be a bottleneck, and let $p(w)$ be the parent of $w$ in the shortest path tree $T$.  If $p(w)$ is a bottleneck, $p(w)$ has smaller level than $w$ and hence precedes $w$ in $L$.  If $p(w)$ is not a bottleneck, let $v$ be the nearest ancestor of p(w) in $T$ that is a bottleneck.  Then $v$ has lower level than $w$, since $w$ dominates all vertices on levels higher than its own.  Since all vertices on the path in $T$ from $v$ to $w$ have true distances at most that of $w$, they are all inserted into $L$ after $v$ but before $w$.  This includes $p(w)$, making the claim true. 
\end{proof}

\subsection{Efficiency of recursive Dijkstra}

Let us bound the time and comparisons used by recursive Dijkstra.  Finding the bottlenecks takes $\OO(m)$ time via a breadth-first search from $s$.  Let $b$ be the number of bottlenecks.  For every marked bottleneck, there are at least two vertices on the next-higher level, both of which are non-bottlenecks.  It follows that the number of marked bottlenecks is at most $(n-b)/2$.  Let $u$ be an unmarked bottleneck, and let $v$ be the bottleneck on the next higher level.  The recursive call $\Dijkstra(u)$ merely inserts $v$ into initially empty $H(u)$, deletes $v$ from $H(u)$, calls $\Dijkstra(v)$, and stops immediately after the recursive call $\Dijkstra(v)$ returns.  Neither the insertion of $v$ into $H(u)$ nor its deletion requires any comparisons.  Thus in counting comparisons we only need to consider the at most $(n-b)/2$ recursive calls of $\Dijkstra$ on marked bottlenecks.  In particular, the heap insertions take $\OO(n)=\OO(m)$ time and $\OO(n-b)$ comparisons.  The latter is $\OO(\log D)$ by~\cref{lem:many-bottlenecks}.

The proof of~\cref{lem:forward-arc-comparisons} shows that recursive Dijkstra scans the vertices in a distance order.  It follows that recursive Dijkstra does at most $F-n+1$ heap \DecreaseKey{} operations, which take $\OO(F-n+1)$ time and comparisons, including each comparison done to determine whether to do a \DecreaseKey{}.

We conclude that the algorithm takes $\OO(m)$ time and does $\OO(F-n+1+\log D)$ comparisons, not including the time and comparisons for heap \DeleteMin{} operations and for searches of $L$ to insert non-bottlenecks.  The next two lemmas bound the time taken and comparisons done by the \DeleteMin{} operations and the searches of $L$, respectively.

\begin{lemma}\label{lem:recursive-Dijkstra-heap-deletions}
The amortized time and number of comparisons done by the \DeleteMin{} operations is $\OO(\log D(G))$.   
\end{lemma}
\begin{proof}
Consider the search tree $T$ generated by a run of recursive Dijkstra.  For each marked bottleneck $v$, let $T(v)$ be the subtree of $T$ containing $v$ and all its descendants connected to it by a path none of whose vertices other than the first and last is a bottleneck.  The number of topological orders of $T$ is at least the product of the number of topological orders of the $T(v)$.  The proof of~\cref{lem:working-set-log-T} applies to the run of $\Dijkstra(v)$ to show that the sum of the logarithms of the working-set sizes of the vertices deleted from $H(v)$ is bounded by a constant times the logarithm of the number of topological orders of $T(v)$.  The number of \DeleteMin{} operations during the runs of $\Dijkstra(v)$ on marked bottlenecks is at most the number of non-bottlenecks plus the number of marked bottlenecks, which is at most $3(n-b)/2$.  Summing over all $v$ and applying~\cref{lem:tree-top-dist} and~\cref{lem:many-bottlenecks} shows that the amortized time and number of comparisons done by the \DeleteMin{} operations is $\OO(\log D(G))$.        
\end{proof}

\begin{lemma}\label{lem:recursive-Dijkstra-search-time}
The amortized time and number of comparisons required for all the searches and insertions in $L$ is $\OO(\log D(G))$.
\end{lemma}
\begin{proof}
Let $v_1=s,v_2,\dots,v_n$ be the vertices in the order they are scanned.  For $1 \leq i \leq n$ let $[a_i,b_i=i]$ be the interval of integers such that $a_1=1$ and, for $i>1$, $a_i=j+1$ where $p(v_i)=v_j$; that is, $a_i$ is one plus the index of $p(v_i)$.  Given that $L$ is represented by a finger search tree, the time and number of comparisons to insert $v_i$ into $L$ is $\O(1+\log(b_i-a_i+1)$.   Let $I$ be the DAG associated with these intervals.  By~\cref{lem:interval-bound}, $\sum_{i=1}^n \log (b_i-a_i+1) = \OO(\log T(I))$, where $T(I)$ is the number of topological orders of $I$.  Let $T$ be the search tree generated by the run of recursive Dijkstra.  Since $p(v)$ is scanned before $v$ for all $v\neq s$, any topological order of $I$ gives a distinct topological order of $T$ by mapping each interval $[a_i,b_i]$ to $v_i$.  It follows that $\sum_{i=1}^n \log (b_i-a_i+1) = \OO(\log D(T) = \OO(\log D(G)$.  Combining this with~\cref{lem:many-bottlenecks} gives the lemma.      
\end{proof}

\begin{remark}
The use of a finger search tree
to support efficient insertions, and the proof of~\cref{lem:recursive-Dijkstra-search-time}, were used previously by~\citet{vanderhoog2024} in a simple algorithm for sorting given the outcomes of a set of pre-existing comparisons.
\end{remark}

\cref{lem:recursive-Dijkstra-heap-deletions} and~\cref{lem:recursive-Dijkstra-search-time} combine to give our desired result:

\begin{theorem}
Recursive Dijkstra runs in time and number of comparisons optimal to within constant factors.
\end{theorem}

\section{A Heap with the Working-Set Bound}\label{sec:heap}

Our final task is to design a heap implementation that has the working-set bound.  Specifically, we need a heap that supports each heap operation except \Meld{} and \DeleteMin{} in $\OO(1)$ amortized time, and that supports each \DeleteMin{} of an item $x$ in $\OO(\log W(x))$ amortized time, where $W(x)$ is the working-set size of $x$.  Since an item cannot be deleted without first being inserted, we can charge $\OO(1)$ of the time spent on a \DeleteMin{} to the insertion of the item that is deleted.  Thus it suffices to obtain an amortized bound of $\OO(1+\log W(x))$ for a \DeleteMin{} of $x$.

There are several known heap implementations that have the working-set bound if \DecreaseKey{} and \Meld{} are not supported operations~\cite{iacono2000improved,elmasry,elmasry_farzan_iacono}.  But we need to support \DecreaseKey{} in $\OO(1)$ time, so we need a new implementation.

\subsection{High-level description}

Our starting point is any heap implementation that has the efficiency of Fibonacci heaps: $\OO(1)$ amortized time for each operation except \DeleteMin{} and $\OO(\log n)$ time for a \DeleteMin{} on an $n$-item heap.  Other examples of such heaps are hollow heaps~\cite{journals/talg/HansenKTZ17} and rank-pairing heaps~\cite{journals/siamcomp/HaeuplerST11}.  We generically call such a heap a \emph{fast heap}.  Our construction is black-box: It can use any fast heap; we do not need to know its internal workings.  We \emph{do} require the fast heap to support 
melds in $\O(1)$ amortized time, however, even though the heap we design does not.

We call our heap an \emph{outer heap}. Our idea is to maintain the set of items in the outer heap very roughly in insertion order, latest first.  To do this we distribute the items in the outer heap among several fast heaps, each of which we call an \emph{inner heap}.  We maintain a list $H_1, H_2,\dots H_k$ of the inner heaps, with the property that every item in $H_i$ was inserted after every item in $H_j$, for any $i<j$.  This property implies that for any item $x$ in $H_i$, the working set of $x$ contains all items in heaps $H_1 ,H_2,\ldots, H_{i - 1}$.  Suppose item $x$ is deleted from inner heap $H_i$.  If, at some time after $x$ is inserted into $H_i$, the size of $H_{i-1}$ is large enough compared to the size of $H_i$ when $x$ is deleted from $H_i$, then the time to delete $x$ will be within the working-set bound.  Working out the details of this idea gives us our desired heap.  As we shall see, heap $H_i$ can grow to a size doubly exponential in $i$, so the number of inner heaps is at most doubly logarithmic in the total number of outer heap insertions.

Our challenge is to control the size of each inner heap.  We do this using melding.  An insertion proceeds in three steps.  First, it creates a new inner heap $H_0$ containing the new item.  Next, it finds the smallest index $j$ such that $H_j$ and $H_{j+1}$ together contain few enough items, and it replaces $H_{j+1}$ by the meld of $H_j$ and the old $H_{j+1}$.  Finally, it reindexes the heaps $H_0, H_1,\dots H_{j-1}$ by increasing the index of each one by $1$, so that $H_i$ becomes $H_{i+1}$ for $0\leq i < j$.   If there is \emph{no} $j$ such that $H_j$ and $H_{j+1}$ together contain few enough items, it does not do a meld but merely reindexes all the inner heaps by increasing the index of each one by $1$.

To complete the specification of insertion, we need a definition of ``few enough."  We define ``few enough" to be ``at most $2^{2^{j+1}}$."

Let $|H|$ denote the size (number of items) in heap $H$.  We implement all the heap operations on an outer heap as follows:

\begin{itemize}
\item[] $\MakeHeap()$: Create and return a new list of inner heaps containing one empty inner heap~$H_1$.

\item[] $\FindMin(H)$: Find an inner heap $H_i$ in $H$ containing an item of minimum key, do a \FindMin{} on $H_i$, and return the item it returns.

\item[] $\DeleteMin(H)$: Find the inner heap $H_i$ in $H$ containing the item returned by $\FindMin(H)$.  Do a \DeleteMin{} on $H_i$ and return the item it returns.

\item[] $\DecreaseKey(x, k, H)$: Find the inner heap $H_i$ in $H$ containing item $x$.  Do a \DecreaseKey{} on $x$ in $H_i$.

\item[] $\Insert(x, H)$: Create a new one-item inner heap $H_0$ containing item $x$.  If there is a $j$ such that $|H_j|+|H_{j+1}| \leq 2^{2^{j+1}}$, choose the minimum such $j$, replace $H_{j+1}$ by the meld of $H_j$ and $H_{j+1}$, and for each $i$ such that $0 \leq i < j$, replace the index of inner heap $H_i$ by $i+1$.  If there is no such $j$, merely replace the index of \emph{every} inner heap $H_i$ by $i+1$.
\end{itemize}

Let us fill in a few more details of the implementation.  Insertion needs access to the heap sizes in order to do the melding test.  The implementation explicitly maintains the size of each inner heap.  This takes $\OO(1)$ time per \Insert{}, \DeleteMin{}, and \Meld{} on an inner heap or heaps.  Decrease-key needs to be able to directly access the location of each heap item in the inner heap containing it.  If the inner heaps are exogenous (the heap nodes are distinct from the items they contain), the implementation stores with each item a pointer to the inner heap node containing it.  If the inner heaps are endogenous (the items themselves are the heap nodes), these pointers are unnecessary.  For a discussion of exogenous versus endogenous data structures, see~\cite{Tarjan1983}. 

Two tasks still require implementation.  First, \DecreaseKey{} needs a way to find the inner heap containing a given item (not just its location within the heap).  To support such queries we maintain a representation of the partition of items among the inner heaps.  Maintaining such a partition is just the classic \emph{disjoint set union} or \emph{union-find} problem, and indeed is a very special case of it.  Second, the \FindMin{} and \DeleteMin{} operations need a way to find an inner heap containing an item of minimum key.  In~\cref{sec:heap-partition} we reduce this task to a tiny instance of another classic data structure problem, that of maintaining a dynamic bit vector subject to queries of the form ``given a bit index, find the closest smaller (larger) index of a $1$ bit."  This problem is sometimes called the \emph{union-split-find} problem.  In our application, there is one bit $b_i$ in the bit vector for each inner heap $H_i$.  Each $1$-bit indicates a suffix-minimum.

We use known solutions to these two problems. These solutions support the needed operations in $\OO(1)$ amortized time.  Before filling in the details, we prove that our heap implementation has the desired efficiency, assuming that we can provide the missing parts.

\subsection{Efficiency}

Insertion is designed to maintain three invariants: (i) All items in an inner heap of smaller index are inserted after all items in an inner heap of larger index; (ii) at all times, $|H_i|$ is bounded from above by a suitable function of $i$; and (iii) if $H_i$ for some $i>0$ changes during an insertion (as the result of a meld or a reindexing), then at the beginning of the insertion (after creation of the new $H_0)$, $|H_{i-1}|$ is bounded from below by a suitable function of $i$.  Together these invariants imply that the amortized time of \DeleteMin{} operations is within the working-set bound.  

\begin{lemma}\label{lem:insertion-order}
If $i < j$, each item in $H_i$ is inserted after each item in $H_j$.   
\end{lemma}
\begin{proof}
The lemma is immediate by induction on time, since each meld and each reindexing during an insertion preserves it, as do the other heap operations.  
\end{proof}

\begin{lemma}\label{lem:heap-size-upper-bound}
At all times, $|H_i| \leq 2^{2^i}$.    
\end{lemma}
\begin{proof}
The lemma is immediate by induction on time, since each meld and each reindexing during an insertion preserves it, including a reindexing of $H_0$ as $H_1$, as do the other heap operations.
\end{proof}

\begin{lemma}\label{lem:inner-heaps}
An outer heap consists of at most $1+\log\log n$ inner heaps, where $n$ is the total number of insertions into the outer heap.
\end{lemma}
\begin{proof}
Let $j$ be the largest index of an inner heap.  After the insertion that gives this inner heap index $j$, the outer heap contains more than $2^{2^{j-1}}$ items, since this insertion does not meld inner heaps $H_{j-2}$ and $H_{j-1}$.  Hence $j \leq 1+\log\log n$.  
\end{proof}

\begin{lemma}\label{lem:heap-size-lower-bound}
Suppose $i>0$ and $H_i$ changes during an insertion.  Then at the beginning of the insertion (after creation of $H_0),$ $|H_{i-1}| > 2^{2^{i-1}}-2^{2^{i-2}}$. 
\end{lemma}
\begin{proof}
Suppose $i=1$.  The new heap $H_0$ created by the insertion contains one item.  Since $1 > 2^{2^0}-2^{2^{-1}}$, the lemma is true for $i=1$.

Suppose $i>1$.  Identify the heaps by their index at the beginning of the insertion, after creation of $H_0$.  Since $i > 1$ and the insertion changes $H_i$, it does not meld the $H_{i-1}$ and $H_{i-2}$.  Hence the $H_{i-2}$ and $H_{i-1}$ together contain more than $2^{2^{i-1}}$ items.  By~\cref{lem:heap-size-upper-bound}, $H_{i-2}$ contains at most $2^{2^{i-2}}$ items.  It follows that $H_{i-1}$ contains at least $2^{2^{i-1}}-2^{2^{i-2}}$ items, making the lemma true for $i>1$.
\end{proof}

\begin{corollary}\label{cor:working-set-size}
If $i>1$, each item in $H_i$ has working-set size greater than $2^{2^{i-2}}-2^{2^{i-3}}$. 
\end{corollary}
\begin{proof}
The only time $H_i$ can acquire new items is if it changes during an insertion.
At the beginning of each such insertion (after creation of $H_0$), $|H_{i-2}| > 2^{2^{i-2}}-2^{2^{i-3}}$ by~\cref{lem:heap-size-lower-bound}.  The insertion reindexes $H_{i-2}$ to be $H_{i-1}$.  By \cref{lem:insertion-order}, all items in $H_{i-1}$ at the end of the insertion are in the working set of all items in $H_i$ at the end of the insertion.  
\end{proof}

\begin{theorem}\label{thm:heap-efficiency}
If finding the inner heap containing a given item and finding an inner heap containing an item of minimum key each take $\OO(1)$ amortized time, then an outer heap has the working-set bound.
\end{theorem}

\begin{proof}
A \MakeHeap{} takes $\OO(1)$ time.  A \FindMin{} takes $\OO(1)$ time by the hypothesis of the theorem.  A \DecreaseKey{} takes $\OO(1)$ time plus the time to find the inner heap containing the item whose key changes plus the time to do a \DecreaseKey{} operation on this inner heap.  The amortized time to do the find is $\OO(1)$ by the hypothesis of the theorem.  The amortized time to do the \DecreaseKey{} in the inner heap is $\OO(1)$ because the inner heap is a fast heap.   

Consider a \DeleteMin{} operation.  Finding the inner heap containing the item to be deleted takes $\OO(1)$ amortized time by the hypothesis of the theorem.  Suppose the item to be deleted is in inner heap $H_i$.  By \cref{lem:heap-size-upper-bound}, $H_i$ contains at most $2^{2^i}$ items just before the deletion.  If $i =1$, the deletion takes $\OO(1)$ time, which is within the working-set bound.  If $i>1$, the working-set size of the deleted item is greater than $2^{2^{i-2}}-2^{2^{i-3}}\geq 2^{2^{i-3}}$ by \cref{cor:working-set-size}.  Since $\log (2^{2^{i-3}}) = 2^{i-3} = 2^i/8 = (\log 2^{2^i})/8$, the deletion time is within the working-set bound.

Consider an \Insert{} operation.  Let $j\geq 1$ be the highest index of a heap changed by the insertion.  The insertion takes $\OO(j)$ time, which we normalize to $j$ units of time.  Identify the heaps by their indices just before the insertion (after creation of $H_0$).  We charge $1$ unit of time to each heap $H_i$ for $0 \leq i < j$.  This pays for the insertion.  We divide this unit charge equally among the items in $H_i$.  By~\cref{lem:heap-size-lower-bound}, an item in $H_i$ is charged at most $1/(2^{2^{i-1}}-2^{2^{i-2}}) \leq 1/2^{2^{i-2}}$ for the insertion.  Each time an item is charged for an insertion, the insertion increases the index of the heap containing it.  The total charge any item accrues is thus at most $\sum_{i=0}^\infty 1/2^{2^{i-2}}$.  Since this sum converges, each item accrues a total charge of $\OO(1)$.  Hence the amortized time per insertion is $\OO(1)$.    
\end{proof}

\subsection{Maintaining the heap partition}\label{sec:heap-partition}

To complete our implementation, we need data structures that (i) maintain the partition of the items among the inner heaps, so that a \DecreaseKey{} can find the inner heap containing a given item, and (ii) maintain a pointer to an inner heap containing an item of minimum key, so that \FindMin{} and \DeleteMin{} can find this heap.  In this section we describe a data structure for the first task, in the next section one for the second task.

The first task is just the classic disjoint set union problem, sometimes called \emph{union-find}: We wish to maintain an initially empty collection of pairwise disjoint non-empty sets, each set having a unique distinguished element called its \emph{leader}, in a way that supports the following three operations:

\begin{itemize}
    \item[] $\MakeSet(v)$: Create a new set containing one element, $v$, which must be in no other set.  Element~$v$ is the leader of the new set.

    \item[] $\Find(v)$: Return the leader of the set containing element $v$.
     
    \item[] $\Unite(v, w)$: Given two elements $v$ and $w$, if $v$ and $w$ are in the same set, do nothing.  If they are in different sets unite these sets, thereby creating a new set containing all the elements in the two old sets containing $v$ and $w$.  Choose any element in the new set as its leader. Uniting the sets destroys the old sets containing~$v$ and~$w$.
\end{itemize}

We maintain the heap partition using a disjoint set data structure as follows:  For each inner heap we maintain a set of its items.  Each heap stores a pointer to the leader of its set of items, and each set leader stores a pointer to the corresponding heap.  Each \DecreaseKey{} takes one \Find{}; each \Meld{} of two inner heaps during an insertion takes one \Unite{}.  We do not update the disjoint-set data structure to reflect deletions.  (But see~\cref{sec:rebuilding}.)

The classic \emph{compressed tree} data structure for disjoint sets~\cite{journals/jacm/Tarjan75}, uses $\OO(n)$ space to store sets containing a total of $n$ elements, takes $\OO(1)$ time worst-case per make-set, and takes $\OO(\alpha(n, m/n))$ amortized time per \Find{} and \Unite{}, if there are a total of $m$ finds, where $\alpha$ is a functional inverse of Ackermann's function.  The $\alpha$ function is very slow-growing: Indeed, its growth is undetectable for any realistic problem size.  This is the data structure to use in practice.  But one can get truly $\OO(1)$ amortized time per \Find{} in special cases.  In particular, if the set (but not the sequence) of \Unite{} operations is known in advance, the \emph{fixed tree} case of disjoint set union, there is an $\OO(1)$ amortized-time solution in the random-access-machine model of computation~\cite{journals/jcss/GabowT85}.

In our use of a heap to implement Dijkstra's algorithm, each item inserted into the heap is eventually deleted.  This is true for Dijkstra with lookahead as well.  In general, if every item inserted into the heap is eventually deleted, there is a simple version of the compressed tree data structure that has the efficiency we need.  To present it, we need to describe the compressed tree data structure in more detail.  Each set is represented by a rooted tree of its elements, with each element having a pointer to its parent.  The root is the leader of the set.  To \Find{} an element, follow the path of parent pointers to the root.  After finding the root, \emph{compress} the path just traversed, by making the root the parent of every node on the path.  To do a \Unite{}, find the roots of the trees containing the given elements.  If they are different, make one the parent of the other.  The new root becomes the new set leader.

The only flexibility in this implementation is the choice of which root becomes the new root when doing a \Unite{}.  When using this data structure in our heap implementation, we make this choice as follows: When melding two inner heaps $H_j$ and $H_{j+1}$, we unite the trees representing $H_j$ and $H_{j+1}$ by choosing the root of the tree representing $H_{j+1}$ to be the root of the new tree.  We call this linking rule \emph{linking by index}.

\begin{lemma}\label{lem:linking-by-index-bound}
Suppose the compressed tree implementation with linking by index is used to maintain the sets of items of the inner heaps.  Let $x$ be an item that is eventually deleted from inner heap $H_j$.  Then the total time for any number of finds of $x$ while it is in the outer heap is $\OO(1)$ per \Find{} plus $\OO(j)$.  
\end{lemma}
\begin{proof}
When the heap $H_j$ containing $x$ changes as the result of an insertion, $j$ increases.  It follows that while $x$ is in the heap it acquires at most $j$ ancestors in its compressed tree.  Each \Find{} of $x$ takes $\OO(1)$ time plus $\OO(1)$ each time the parent of $x$ changes.  There can be at most $j$ such changes.
\end{proof}

When an item $x$ is deleted from heap $H_j$ such that $j>1$, the logarithm of its working-set size is at least $\log(2^{2^{j-2}}-2^{2^{j-3}})$ by~\cref{cor:working-set-size}.  Thus we can charge the $\OO(j)$ extra time for finds of $x$ to the deletion of $x$ from the heap, giving us an $\OO(1)$ amortized time bound for each \Find{} and \Unite{}, which is what the hypothesis of~\cref{thm:heap-efficiency} requires.

Although this simple solution suffices for our use of a heap in Dijkstra's algorithm, let us consider the general case, in which some items are never deleted from the heap.  For the general case there is a more-complicated $\OO(1)$ amortized-time solution.  We use fixed-tree set union. We give each item inserted into a heap an identifier, with identifiers being the successive natural numbers.  If an item is deleted and later reinserted, we treat it as a new item.  The set of possible unites is $\Unite(i, i+1)$ for $i$ from $1$ to $n-1$, if $n$ is the number of heap insertions.  The Gabow-Tarjan data structure~\cite{journals/jcss/GabowT85} for fixed-tree disjoint set union does what we need.


\subsection{Maintaining the heap minimum}\label{sec:heap-minimum}
Our second and final task is to maintain a pointer to an inner heap containing an item of minimum key.  We shall reduce this problem to one of maintaining a very short dynamic bit vector, which we can do in $\OO(1)$ time per operation using the power of bit vector manipulation allowed in the RAM model.

We store pointers to the inner heaps in an array indexed by heap index.  This supports access to an inner heap in $\OO(1)$ time given its index.  Reindexing during an insertion requires updating this array, but the time required is at most a constant factor times the reindexing time, which is $\OO(1)$ per outer heap insertion by~\cref{thm:heap-efficiency}, and hence within the working-set bound.

We call an inner heap $H_i$ a \emph{suffix minimum} if it is non-empty and the minimum of the keys of its items is less than that of every non-empty inner heap $H_j$ such that $j>i$.  In particular, the highest-index non-empty inner heap is a suffix minimum.  We maintain a bit vector $b$ indexed by heap index such that $b_i=1$ if $H_i$ is a suffix minimum, $b_i=0$ if not.  Vector $b$ has a bit of index $0$ for the temporary inner heap $H_0$ created during an insertion.  Between insertions, $b_0=0$.  In addition to being able to read or write any bit given its index, we need the following two query operations on $b$:

\begin{itemize}

    \item[] $\Next(i)$: Return the smallest $j \geq i$  such that $b_i=1$, or $\Null$ if there is no such $j$.

    \item[] $\Prev(i)$: Return the largest $j \leq i$ such that $b_j=1$, or $\Null$ if there is no such $j$.
     
\end{itemize}

We assume that each bit read, write, or query operation takes $\OO(1)$ time.  (We justify this assumption later in this section.)  We use and maintain $b$ during the outer heap operations as follows:

\begin{itemize}
\item[] $\MakeHeap()$: Create and return a new list of inner heaps containing one empty inner heap~$H_1$.  Initialize an all-zero bit vector $b$ of length at least $1+\log\log n$, where $n$ is the anticipated number of insertions. 

\item[] $\FindMin(H)$: Set $j\gets\Next(1)$.  If $j\neq \Null$, inner heap $H_j$ contains an item of minimum key.  Do a \FindMin{} on $H_j$, and return the item it returns.  If $j=\Null$, the entire outer heap is empty.  Return $\Null$.

\item[] $\DeleteMin(H)$: Set $j\gets\Next(1)$.  If $j\neq\Null$, proceed as follows.  Do a \DeleteMin{} on $H_j$ and save the item returned. Set $k\gets\Next(j+1)$.  If $k\neq \Null$, $H_k$ is the next suffix minimum after $H_j$.  If $k \neq\Null$, do a \FindMin{} on $H_k$.  Starting from $H_j$ and proceeding through inner heaps of smaller index, determine which of these heaps are now suffix minima, and update $b$ accordingly.  To do this, initialize the current minimum key to be $\infty$ if $k=\Null$, or the key of the item returned by the \FindMin{} on $H_k$ if $k\neq\Null$.  Each successive inner heap is a suffix minimum if its minimum key (found by doing a \FindMin{}) is less than the current minimum key; if it is, set its bit in $b$ to $1$ and set the current minimum key to the minimum key of an item in this heap.  If an inner heap is not a suffix minimum, set its bit to $0$ and do not change the current minimum key.  The time spent updating $b$ is $\OO(j)$. After updating $b$, return the saved item of minimum key.    

\item[] $\DecreaseKey(x, k, H)$: Find the inner heap $H_j$ in $H$ containing item $x$ using the set-union data structure of~\cref{sec:heap-partition}.  Do a \DecreaseKey{} on $x$ in $H_j$.  If $x$ does not now have minimum key among the items in $H_j$, this completes the \DecreaseKey{}.  If it does have minimum key, then $H_j$ may now be a suffix minimum, and some or all of the suffix minima $H_i$ with $i<j$ may have become non-suffix minima.  To update $b$ in this case, set $k\gets\Next(j+1)$.  Test whether $H_j$ is now a suffix minimum by comparing the new key of $x$ with that of the key of  $\FindMin(H_k)$.  (If $k=\Null$ then $H_j$ is definitely a suffix minimum.)  Set $b_j$ accordingly.  If $H_j$ is not a suffix minimum, this completes the \DecreaseKey{}.  If it is a suffix minimum, proceed through suffix minima $H_i$ with $i < j$ in decreasing order on $i$ to find those that are no longer suffix minima.  To do this, initialize $i\gets\Prev(j-1)$.  Repeat the following step until $H_i$ is a suffix minimum or $i=\Null$: Test whether $H_i$ is a suffix minimum by comparing the key of $\FindMin(i)$ with that of $x$. If not, set $b_i=0$, and set $i\gets \Prev(i)$.

\item[] $\Insert(x, H)$: Create a new one-item inner heap $H_0$ containing item $x$.  Test whether $H_0$ is a suffix minimum by comparing its key with that of the  inner heap whose index is $\Next(1)$, and set $b_0$ appropriately.  (If $\Next(1)=\Null$, $H_0$ is a suffix minimum.)  If there is a $j$ such that $|H_j|+|H_{j+1}| \leq 2^{2^{j+1}}$, choose the minimum such $j$, replace $H_{j+1}$ by the meld of $H_j$ and $H_{j+1}$, and for each $i$ such that $0 \leq i < j$, replace the index of inner heap $H_i$ by $i+1$.  If there is no such $j$, merely replace the index of \emph{every} inner heap $H_i$ by $i+1$.  Starting from the highest-index heap changed, proceed through heaps of decreasing index, determining which ones are now suffix minima and setting their bits accordingly.  Set $b_0=0$.  This updating is the same as the updating done in a \DeleteMin{}.  The time spent updating $b$ is $\OO(j)$.    
\end{itemize}

It is straightforward to verify that this implementation correctly maintains $b$, and hence that the entire implementation is correct.  The extra time needed to maintain $b$ is also within the working-set bound, as we now prove:

\begin{lemma}
If each bit vector operation takes $\OO(1)$ time, the time spent using and maintaining $b$ during a sequence of heap operations is within the working-set bound.  
\end{lemma}
\begin{proof}
A \FindMin{} does one bit operation.  A \DeleteMin{} does one bit operation to find the inner heap $H_j$ of smallest index that is a suffix minimum and spends $\OO(j)$ time updating $b$.  By~\cref{cor:working-set-size}, if $j>1$ heap $H_j$ has working-set size greater than $2^{2^{j-2}}-2^{2^{j-3}}$.  The logarithm of its working-set size is thus more than enough to cover the time spent updating $b$.  An insertion that changes $j$ inner heaps spends $\OO(j)$ time updating $b$, which is $\OO(1)$ amortized by the proof of~\cref{thm:heap-efficiency}.  Finally, a \DecreaseKey{} spends $\OO(1)$ time plus $\OO(1)$ per bit changed from $1$ to $0$ updating $b$.  We charge the time to flip bits from $1$ to $0$ to the events that flipped them from $0$ to $1$.  With this charging argument, the amortized time spent by a \DecreaseKey{} updating $b$ is $\OO(1)$.       \end{proof}

Now we just need a way to implement the bit operations.  For this we use random access.  The bit vector $b$ can be very short: It only needs to hold $1+\log\log n$ bits, where $n$ is the total number of insertions in the outer heap.  We store $b$ in a single word of computer memory.  Accessing or flipping a bit takes $\OO(1)$ time.  The $\Next$ and $\Prev$ operations can be done with a constant number of appropriate mask and shift operations.  Alternatively, since there are $\OO(\log n)$ possible bit vector values, one can construct tables of the values of the $\Next$ and $\Prev$ functions and use table lookup to do each operation: Each lookup table needs at most $\log n$ words of memory.

This solution requires that we know $n$ in advance, but a very loose upper bound suffices.  We deal with this issue in~\cref{sec:rebuilding}.

We conclude this section with some remarks about the general problem of maintaining a bit vector subject to bit reads, writes, and $\Next$ and $\Prev$ queries.  This problem has been called the \emph{union-split-find} problem, because it is equivalent to maintaining a partition of the interval $[1, k]$ of the natural numbers $1$ through $k$ into sub-intervals, subject to queries that find the sub-interval containing a given number and updates that split an interval into two or combine two adjacent intervals into one. \citet{journals/ipl/Boas77} gave a solution to this problem in which each operation takes $\OO(\log\log k)$ time.  In our case we have a very tiny instance of the problem, which is what allows us to get an $\OO(1)$ time bound.  

\subsection{Heap rebuilding}\label{sec:rebuilding}
The space required by our implementation of an outer heap is $\OO(n)$, where $n$ is the total number of insertions.  If most of the inserted items are deleted, the space used by the data structure can become greater than linear in the number of items currently in the heap.  Furthermore, the bit vector needed to keep track of the inner heap suffix minima (see~\cref{sec:heap-minimum} can become long compared to the current heap size.  Finally, the fixed-tree set union data structure used to maintain the heap partition in the general case in~\cref{sec:heap-partition} and method for doing bit vector operations in~\cref{sec:heap-minimum} both require an upper bound on $n$, although this bound can be very loose.  The standard way to address such issues is to completely rebuild the data structure each time its current size decreases to a suitable constant fraction of its maximum size since its last rebuilding, or the number of insertions done since the last heap rebuilding exceeds a suitable constant times the size of the data structure after the last rebuilding.  In our case, rebuilding takes linear time in the current size of the data structure, so we can use this technique without affecting our amortized time bounds.  If rebuilding is used, we need to maintain a list of the items in the heap in insertion order, so that we can re-insert them in the correct order.  In the application to Dijkstra's algorithm, we know $n$ in advance, so rebuilding is unnecessary.

\section{Remarks}\label{sec:remarks}

We have proved that Dijkstra's algorithm implemented with a sufficiently efficient heap is universally optimal in running time, and that two different extensions of the algorithm are universally optimal in both time and comparisons.  We have also developed a heap with the needed efficiency.

We are optimistic that the concept of universal optimality can be fruitfully applied to other problems.  One that intrigues us is the single-source, single-target version of the distance order problem, in which a target vertex $t$ as well as a source vertex $s$ is given, and the problem is to return a list of vertices containing $s$ and $t$ that is a prefix of some distance order.  If we stop Dijkstra's algorithm as soon as $t$ is deleted from the heap, we obtain a solution to this problem.  To make this algorithm universally optimal for this problem seems to require a heap with a stronger working-set bound that does not count items inserted into the heap but never deleted.  Designing a heap with the needed efficiency that supports decrease-key in $\OO(1)$ amortized time remains an open problem.   

\begin{acks}
Partially funded by the Ministry of Education and Science of Bulgaria's support for INSAIT as part of the Bulgarian National Roadmap for Research Infrastructure. 
BH and RH were supported in part by the European Research Council (ERC) under the European Union's Horizon 2020 Research and Innovation Programme (grant agreement No.~949272).
VR was supported by the European Research Council (ERC) under the European Union's Horizon 2020 Research and Innovation Programme (grant agreement No.~853109). 
RT was partially supported by a gift from Microsoft.
JT and RH were supported by the VILLUM Foundation grants 54451 and 16582.

JT worked on this paper while affiliated with BARC, University of Copenhagen.
RH partially worked on this paper while visiting BARC, University of Copenhagen.
VR worked on this paper while visiting MIT and while affiliated with ETH Zurich.
RT worked on this paper while visiting the Simons Institute for the Theory of Computing and INSAIT.
\end{acks}

\ifconference
\bibliographystyle{IEEEtran}
\bibliography{IEEEabrv,ref}
\else
\printbibliography
\fi

\end{document}